\documentclass[11pt]{article}
\usepackage{amsfonts}
\usepackage{amssymb}
\usepackage{mathrsfs}
\usepackage{graphicx}
\usepackage{amsmath,amsthm,amssymb,amscd}
\usepackage[all]{xy}
\usepackage{subfig}
\usepackage{hyperref}
\usepackage{multirow}

\usepackage[dvips]{color}
\usepackage{amsfonts}
\usepackage{float}
\def\eqa{\begin{align}}
\def\eqae{\end{align}}
\def\eq{\begin{equation}}
\def\eqe{\end{equation}}
\def\be{\begin{equation}}
\def\ee{\end{equation}}
\def\ba{\begin{array}}
\def\ea{\end{array}}
\def\bd{\begin{displaymath}}
\def\ed{\end{displaymath}}

\def\>{\rangle}
\def\<{\langle}

\numberwithin{equation}{section}

\begin{document}

\begin{titlepage}
\begin{flushright}
PUPT 2454 \\
\end{flushright}
\vspace{1cm}

\begin{center}
{\Large \bf Higher-spin massless S-matrices in four-dimensions}\\[1cm]
David A. McGady and Laurentiu Rodina\\
{\it Department of Physics, Princeton University}\\
{\it Princeton, NJ 08544, USA}\\
\today

\end{center}

\abstract{On-shell, analytic S-matrix elements in massless theories are constructed from a finite set of primitive three-point amplitudes, which are fixed by Poincare invariance up to an overall numerical constant. We classify \emph{all} such three-point amplitudes in four-dimensions. Imposing the simplest incarnation of Locality and Unitarity on four-particle amplitudes constructed from these three-particle amplitudes rules out all but an extremely small subset of interactions among higher-spin massless states. Notably, the equivalence principle, and the Weinberg-Witten theorem, are simple corollaries of this principle. Further, no massless states with helicity larger than two may consistently interact with massless gravitons. Chromodynamics, electrodynamics, Yukawa and $\phi^3$-theories are the only marginal and relevant interactions between massless states. Finally, we show that supersymmetry naturally emerges as a consistency condition on four-particle amplitudes involving spin-3/2 states, which must always interact gravitationally.}

\end{titlepage}

\tableofcontents
\newpage

\section{Consistency conditions on massless S-matrices}\label{summary}

Pioneering work by Weinberg showed that simultaneously imposing Lorentz-invariance and unitarity, while coupling a hard scattering process to photons, necessitates \emph{both} charge conservation and the Maxwell equations~\cite{SoftQED}. Similarly, he showed the same holds for gravity: imposing Lorentz-invariance and unitarity on hard scattering processes coupled to gravitons implies both the equivalence principle and the Einstein equations~\cite{SoftGR}. Weinberg's theorems were then extended to fermions in \cite{susy.ext}, where it was shown that spin-3/2 particles lead to supersymmetry. In the case of higher spin theories \cite{hs1}, which are closely related to string theory \cite{string}, it was shown that unitarity and locality impose severe restrictions, and many no-go theorems were established \cite{hs.constraint}\cite{WW}, more recently including in CFT's \cite{cft.hs1}\cite{cft.hs2}.

The goal of this paper is to systematize and extend prior the analysis of Refs.~\cite{BC} and~\cite{ST} of the leading-order interactions between any set of massless states in four dimensions, within the context of the \emph{on-shell} perturbative S-matrix. In short, our results are: (1) a new classification of three-particle amplitudes in constructible massless S-matrices, (2) ruling out all S-matrices built from three-point amplitudes with $\sum_{i = 1}^3 h_i = 0$ (other than $\phi^3$-theory), (3) a new on-shell proof of the uniqueness of interacting gravitons and gluons, (4) development of a new test on four-particle S-matrices, and (5) showing how supersymmetry naturally emerges from consistency constraints on certain four-particle amplitudes which include spin-3/2 particles. 

Massless vectors (gluons and photons) and tensors (gravitons) are naturally described via on-shell methods~\cite{BC}-\cite{UCLAnSLAC}. On- and off-shell descriptions of these massless higher-spin states are qualitatively different: on-shell they have only transverse polarization states, while off-shell all polarization states may be accessed. Local field theory descriptions necessarily introduce these unphysical, longitudinal, polarization states. They must be removed through introducing extra constraints which ``gauge them away''. Understanding consistency conditions on the interactions of massless gravitons and gluons/photons, should therefore stand to benefit from moving more-and-more on-shell, where gauge-invariance is automatic.

``Gauge anomalies'' provide a recent example~\cite{tHooft}\cite{GS}. What is called a ``gauge-anomaly'' in off-shell formulations, on-shell is simply a tension between parity-violation and locality. Rational terms in parity-violating loop amplitudes either do not have local descriptions, or require the Green-Schwarz two-form to restore unitarity to the S-matrix~\cite{Anomalies}.

Along these lines, there are recent, beautiful, papers by Benincasa and Cachazo~\cite{BC}, and by Schuster and Toro~\cite{ST} putting these consistency conditions more on-shell. Ref.~\cite{BC} explored the constraints imposed on a four-particle S-matrix through demanding consistent on-shell BCFW-factorization in various channels on the coupling constants in a given theory~\cite{BCFW}.\footnote{This consistency condition is further explored, for instance, in ref.~\cite{Benincasa}.}  Four-particle tests based on BCFW have been used to show the inconsistency of some higher spin interactions in Refs. \cite{hs.test1}\cite{hs.test2}, where it was suggested that non-local objects must be included in order to provide consistent theories (see also \cite{nonlocal}). 

The analysis of Ref.~\cite{BC} hinged upon the existence of a ``valid'' BCF-shift, 
\begin{align}
\nonumber A_n(z) = A_n(\{p_i+qz, p_j-qz\}) \,\, , \, {\rm with} \,\,\, p_i  \cdot q = p_j \cdot q = 0  \,\,\, {\rm and} \,\,\, q \cdot q = 0,
\end{align}
of the amplitude that does not have a pole at infinity. If $A(z)$ does not have a large-$z$ pole, then its physical value, $A(0)$, is a sum over its residues at the finite-$z$ poles. These finite-$z$ poles are factorization channels; their residues are themselves products of lower-point \emph{on-shell} amplitudes: an on-shell construction of the whole S-matrix~\cite{BCF}\cite{allLoop}\cite{CSW}. However, then extant existence proofs for such shifts resorted to local field theory methods~\cite{BCFW}\cite{SpinLorentz}\cite{Risager}.

Hence, ref.~\cite{ST} relaxes this assumption, through imposing a generalized notion of unitarity, which they refer to as ``complex factorization''. The consequences of these consistency conditions are powerful. For example, they uniquely fix (1) the equivalence of gravitational couplings to all matter, (2) decoupling of multiple species of gravitons within S-matrix elements, and (3) the Lie Algebraic structure of spin-1 interactions.

Our paper is organized as follows. We begin in section~\ref{basics} by developing a useful classification of all on-shell massless three-point amplitudes. We here pause to make contact with standard terminology for ``relevant'', ``marginal'' and ``irrelevant'' operators in off-shell formulations of Field Theory, and to review basic tools of the on-shell S-matrix. 

This is applied in section~\ref{PoleCount}, first, to constructible four-particle amplitudes in these theories. Locality and unitarity sharply constrain the analytic structure of scattering amplitudes. Specifically, four-point amplitudes cannot have more than three poles. Simple pole-counting, using the classification system in section~\ref{basics}, rules out all lower-spin theories, save $\phi^3$-theory, (S)YM, and GR/SUGRA--and one pathological example, containing the interaction vertex $A_3(\frac{1}{2},-\frac{1}{2},0)$. In section~\ref{UniqueLaws}, we show that the gluons can only consistently interact via YM, GR and the higher-spin amplitude $A_3(1,1,1)$. Similarly, gravitons can only interact via GR and the higher-spin amplitude $A_3(2,2,2)$. Gravitons and gluons are unique, and cannot couple to higher-spin states. These two sections strongly constrain the list of possible interacting high-spin theories, in accordance with existing no-go theorems.

Utilizing the information in sections~\ref{PoleCount} and ~\ref{UniqueLaws}, in section~\ref{niceshift} we derive and apply a systematic four-particle test, originally discussed in Ref.~\cite{PITP}. This test independently demonstrates classic results known about S-matrices of massless states, such as the equivalence principle, the impossibility of coupling gravitons to massless states with $s > 2$ ~\cite{gr.decouple}, decoupling of multiple spin-2 species, and the Lie Algebraic structure of vector self-interactions.

Knowing the equivalence principle, in section~\ref{SUSY} we then study the consistency conditions of S-matrices involving massless spin-3/2 states. From our experience with supersymmetry, we should expect that conserved fermionic currents correspond to massless spin-3/2 states. In other words, we expect that a theory which interacts with massless spin-3/2 particles should be supersymmetric. 

Supersymmetry manifests itself through requiring all poles within four-point amplitudes have consistent interpretations. The number of poles is fixed, mandated by locality and unitarity and the mass-dimension of the leading-order interactions. Invariably, for S-matrices involving external massless spin-3/2 particles, at least one of these poles begs for inclusion of a new particle into the spectrum, as a propagating internal state on the associated factorization channel. For these S-matrices to be consistent, they \emph{require} both gravitons to be present in the spectrum \cite{sugra.c} and supersymmetry. We close with future directions in section~\ref{Conclusion}.

\section{Basics of on-shell methods in four-dimensions}\label{basics}

In this section, we briefly review three major facets of modern treatments of massless S-matrices: the spinor-helicity formalism, kinematic structure of three-point amplitudes in these theories, and the notion of constructibility. The main message is three-fold:
\begin{itemize}
\item The kinematic dependence of three-point on-shell amplitudes is uniquely fixed by Poincare invariance (and is best described with the spinor-helicity formalism).
\item On-shell construction methods, such as BCFW-recursion, allow one to recursively build up the entire S-matrix from these on-shell three-point building blocks. Amplitudes constructed this way are trivially ``gauge-invariant''. There are no gauges.
\item Any pole in a local and unitary scattering amplitude must both (a) be a simple pole in a kinematical invariant, e.g. $1/K^2$, and (b) have a corresponding residue with a direct interpretation as a factorization channel of the amplitude into two sub-amplitudes.    
\end{itemize}

\subsection{Massless asymptotic states and the spinor-helicity formalism}

In a given theory, scattering amplitudes can only be functions of the asymptotic scattering states. Relatively few pieces of information are needed to fully characterize an asymptotic state: momentum, spin, and charge/species information. Spinor-helicity variables automatically and fully encode both momentum and spin information for massless states in four-dimensions.

Four-dimensional Lorentz vectors map uniquely into bi-spinors, and vice versa (the mapping is bijective): $p_{\alpha \dot{\alpha}} = p_{\mu} \sigma^{\mu}_{\alpha \dot{\alpha}}$. Determinants of on-shell momentum bi-spinors are proportional to $m^2$. Bi-spinors of massless particles thus have rank-1, and must factorize into a product of a left-handed and a right-handed Weyl spinor: $p^2 = 0 \Rightarrow p^{\alpha \dot{\alpha}} = \lambda^{\alpha} \tilde{\lambda}^{\dot{\alpha}}$. 

These two Weyl spinors $\lambda$ and $\tilde{\lambda}$ are the spinor-helicity variables, and are uniquely fixed by their corresponding null-momentum, $p$, up to rescalings by the complex parameter $z$: $(\lambda, \tilde{\lambda}) \rightarrow (z\lambda, \tilde{\lambda}/z)$. Further, they transform in the (1/2,0) and (0,1/2) representations of the Lorentz group. Dot products of null momenta have the simple form, $p_i \cdot p_j = \langle i j \rangle [ j i ]$, where the inner-product of the (complex) LH-spinor-helicity variables, is $\langle A B \rangle \equiv \lambda^A_{\alpha} \lambda^B_{\beta} \epsilon^{\alpha \beta} $, and the contraction of the RH- Weyl spinors is $[ A B ] \equiv \tilde{\lambda}^A_{\dot{\alpha}} \tilde{\lambda}^B_{\dot{\beta}} \epsilon^{\dot{\alpha} \dot{\beta}}$.

A good deal of the power of the spinor-helicity formalism derives from the dissociation between the left-handed and right-handed degrees of freedom. Real null-momenta are defined by the relation,
\eq
\tilde{\lambda} = \bar{\lambda}
\eqe
between the two Weyl-spinors. Complex momenta are not similarly bound: the left-handed and right-handed Weyl-spinors need not be related for complex momentum. For this reason, they can be independently deformed by complex parameters; this efficiently probes the analytic properties of on-shell amplitudes that depend on these variables. From here on out, we refer to the left-handed Weyl-spinors, i.e. the $\lambda$s, as \emph{holomorphic} variables; right-handed Weyl-spinors, i.e. the $\tilde{\lambda}$s are referred to as \emph{anti-holomorphic} variables. Similarly, \emph{holomorphic} spinor-brackets and \emph{anti-holomorphic} spinor-brackets refer to $\langle \lambda, \chi \rangle$- and $[ \tilde{\lambda}, \tilde{\chi}]$-contractions.

Identifying the ambiguity $(\lambda, \tilde{\lambda}) \rightarrow (z\lambda, \tilde{\lambda}/z)$ with little-group (i.e. helicity) rotations, $(\lambda, \bar{\lambda}) \rightarrow (e^{-i \theta/2}\lambda, e^{i \theta/2}\bar{\lambda})$, allows one to use the spinor-helicity variables to express not only the momenta of external states in a scattering process, but also their spin (helicity). In other words, the spinor-helicity variables encode all of the data needed to characterize massless asymptotic states, save species information.

\subsection{Three-point amplitudes}\label{3point}

Scattering processes involving three massless on-shell states have no non-trivial kinematical invariants. At higher-points, complicated functions of kinematical invariants exist that allow rich perturbative structure at loop level. These invariants are absent at three-points. Poincare invariance, up to coupling constants, thus uniquely and totally fixes the kinematical structure of all three-point amplitudes for on-shell massless states. 

The standard approach to solving for the three-point amplitudes (see for example Ref. \cite{BC}) involves first writing a general amplitude as:
\begin{align}
\label{bothforms}
A_3=A_3^{\left(\lambda\right)}(\langle 12\rangle,\langle 23\rangle,\langle 31\rangle)+A_3^{\left(\tilde{\lambda}\right)}([12],[23],[31])
\end{align}
where $(\lambda)$ denotes exclusive dependence on holomorphic spinors, and $(\tilde{\lambda})$ denotes the same for anti-holomorphic spinors. Imposing momentum conservation forces $[12] = [23] = [31] = 0$, and/or $\langle 12 \rangle = \langle 23 \rangle = \langle 31 \rangle = 0$. Typically, only one of the two functions in Eq. (\ref{bothforms}) is smooth in this limit, and is thus selected as the physical one, while the other is discarded.

Explicitly, in these cases, the amplitudes become:
\begin{align}
&A_3(1^{h_1}_a,2^{h_2}_b,3^{h_3}_c) = g^{-}_{abc}
\langle 12 \rangle^{h_3 - h_1 - h_2} \langle 23 \rangle^{h_1 - h_2 - h_3} \langle 31 \rangle^{h_2 - h_3 - h_1} \, , \, {\rm for} \,\sum_{i=1}^3 h_i < 0 \, ,  \,  \nonumber \\ 
&A_3(1^{h_1}_a,2^{h_2}_b,3^{h_3}_c) = g^{+}_{abc}
[ 12 ]^{h_1 + h_2 - h_3} [ 23 ]^{h_2 + h_3 - h_1} [ 31 ]^{h_3 + h_1 - h_2} \,\,\,\,\, , \, {\rm for} \,\sum_{i=1}^3 h_i > 0 \, , \label{3ptA3}
\end{align}
where $g^{\pm}_{abc}$ is the species dependent coupling constant. 

However, this approach leads to ambiguities in the $\sum_{i=1}^3 h_i = 0$ case. Consider for example a three-point interaction between two opposite-helicity fermions and a scalar. Equation (\ref{bothforms}) reads in this case:
\begin{align}
A_3\left(1^0, 2^{-\frac{1}{2}},3^{\frac{1}{2}}\right)=g^-\frac{\langle 12\rangle}{\langle 13\rangle}+g^+\frac{[13]}{[12]}
\end{align}
 Imposing momentum conservation, for example by setting $\langle 12 \rangle = \langle 23 \rangle = \langle 31 \rangle = 0$, is clearly ill-defined\footnote{Attempting to impose momentum conservation by a well-defined limit leads to other inconsistencies as well, for example with the helicity operator.}. Because of this ambiguity, $\sum_{i=1}^3 h_i = 0$ amplitudes have generally been ignored in most of the on-shell literature. However, the ambiguity is only superficial. 

The inconsistencies arise because we first find the most general eigenfunction of the helicity operator, ie. Eq. (\ref{bothforms}), and only after that do we impose momentum conservation. However, this order of operations is arbitrary. Since we always only deal with on-shell amplitudes, we can simply first fix for example $\langle 12 \rangle = \langle 23 \rangle = \langle 31 \rangle = 0$, and then look for solutions which are functions only of $\tilde{\lambda}$s. In this case, the amplitudes are perfectly well defined as:
\begin{align}
A_3&=g^-_{abc} f^-(\lambda_i), \quad \textrm{when $[12] = [23] = [31] = 0$}\\
&\textrm{and}\nonumber\\
A_3&=g^+_{abc} f^+(\tilde{\lambda}_i),\quad \textrm{when $\langle 12 \rangle = \langle 23 \rangle = \langle 31 \rangle = 0$}
 \end{align}
Ultimately, it will in fact turn out that none of these amplitudes are consistent with locality and unitarity, but this approach clears any ambiguities related to $\sum_{i=1}^3 h_i = 0$ amplitudes.

Before moving on, we pause to consider the role of parity in the on-shell formalism. Parity conjugation swaps the left-handed and right-handed $SU(2)$s that define the (double-cover) of the four-dimensional Lorentz-group. As such, parity swaps the left-handed Weyl-spinors with the right-handed Weyl-spinors, $(1/2,0)\leftrightarrow (0,1/2)$. Therefore, within the spinor-helicity formalism, in the context of Eq.~\eqref{3ptA3}, 
\begin{align}
g^{-}_{abc} = g^{+}_{abc}& \Longleftrightarrow \, {\rm Parity - conserving \,\, interactions} \, \, , \, \, {\rm and} \\ 
g^{-}_{abc} = - g^{+}_{abc} &\Longleftrightarrow \, {\rm Parity - violating \,\, interactions} \, . \nonumber
\end{align}
Further, as we associate the right-handed Weyl-spinors, i.e. the $\lambda$s, with holomorphic degrees of freedom and the left-handed Weyl-spinors, i.e. the $\tilde{\lambda}$s, with anti-holomorphic degrees of freedom, we see that parity-conjugation swaps the holomorphic and anti-holomorphic variables. In other words, parity- and complex- conjugation are one-and-the-same. The conjugate of a given three-point amplitude is the same amplitude with all helicities flipped: the ``conjugate'' of $A_3(1^{+h_1},2^{+h_2},3^{+h_3})$ is $A_3(1^{-h_1},2^{-h_2},3^{-h_3})$.

We will find it useful to classify all such three-particle amplitudes by two numbers:
\begin{align}
 A = \bigg| \sum_{i = 1}^3 h_i \bigg| \, , \,\, H = {\rm max}\bigg\{ |h_1|, |h_2|, |h_3| \bigg\} \, .  \label{HA}
\end{align}
Comparing the relevant operator in $\phi^3$-theory to its corresponding primitive three-point amplitude, we infer that three-point amplitudes with $A = 0$ correspond to \emph{relevant} operators. Similarly, QCD's $A = 1$ three-point amplitude corresponds to \emph{marginal} operators; GR has $A = 2$, and interacts via irrelevant, $1/M_{pl}$ suppressed, operators. 

\subsection{Four points and higher: Unitarity, Locality, and Constructibility}

There are several, complimentary, ways to build up the full S-matrix of a theory, given its fundamental interactions. Conventionally, this is through Feynman diagrams, the work-horse of any perturbative analysis of a given field theory. However, this description of massless vector- (and higher-spin-) scattering via local interaction Lagrangians necessarily introduces unphysical, longitudinal, modes into intermediate expressions~\cite{SimplestQFT}\cite{PITP}. To project out these unphysical degrees of freedom, one must impose the gauge conditions.

On the other hand, recent developments have elucidated methods to obtain the full S-matrix, while keeping \emph{all} states involved on-shell (and physical) throughout the calculation~\cite{BC}\cite{ST}\cite{UCLAnSLAC}\cite{BCFW}\cite{allLoop}\cite{Construct}. We refer to these methods, loosely speaking, as ``constructive''.  Crucially, because all states are on-shell, all degrees of freedom are manifest, thus: \emph{amplitudes that are directly constructed through on-shell methods are automatically gauge-invariant.} This simple fact dramatically increases both (a) the computational simplicity of calculations of scattering amplitudes, and (b) the physical transparency of the final results.

The cost is that amplitudes sewn together from on-shell, delocalized, asymptotic states do not appear to be manifestly local. Specifically, at the level of the amplitude, locality is reflected in the pole-structure of the amplitude. Scattering amplitudes in local theories have exclusively propagator-like, $\sim 1/K^2$, poles ($K = \sum_i p_i$ is a sum of external null momenta). Non-local poles correspond to higher-order poles, i.e. $1/(K^2)^{4}$, and/or poles of the form, $1/\langle i | K | j]$, where $K$ is a sum of external momenta.\footnote{Indeed, individual terms within gluon amplitudes generated by, for instance, BCFW-recursion\cite{BCF}\cite{BCFW} contain ``non-local'' poles, specifically of this second type, $\sim1/\langle i | K | j]$. These non-local poles, however, always cancel in the total sum, and the final expression is manifestly local~\cite{SimplestQFT}\cite{allLoop}.} An on-shell S-matrix is local if its only kinematical poles are of the form $1/(\sum_i p_i)^2$.

Unitarity, as well, has a slightly different incarnation in the on-shell S-matrix. In its simplest guise, unitarity is simply the dual requirement that (a) the residue on each and every pole in an amplitude \emph{must} have an interpretation as a physical factorization channel,
\begin{align}
A^{(n)} \rightarrow \frac{1}{K^2} \, A_L^{(n-m+1)} \times A_R^{(m+1)} \ , \label{Unitarity1}
\end{align}
and (b) that any individual factorization channel, if it is a legitimate bridge between known lower-point amplitudes in the theory, must be a residue of a fully legitimate amplitude with the same external states in the theory. For example, given a factorization channel of the form $A_3(1^{-2},2^{-2},P_{12}^{+2}) \frac{1}{s_{12}}  A_3(P_{12}^{-2},3^{+2},4^{+2})$, within a theory \emph{constructed} from the three-point amplitude $A_3(+2,-2,+2)$ and its parity-conjugate, \emph{then} this must be a factorization channel of the four-point amplitude $A_4(1^{-2},2^{-2},3^{+2},4^{+2})$.

Poincare invariance uniquely fixes the three-particle S-matrix in a theory, up to coupling constants. Constructive methods, such as the BCFW recursion relations, use these fixed forms for the three-point amplitudes as input to build up the entire S-matrix, without making reference to Feynman diagrams~\cite{BCFW}\cite{allLoop}. Basic symmetry considerations, residue theorems, and judicious application of tree-level/single-particle unitarity, fix the entire S-matrix!\footnote{Invocations of ``unitarity'' in this paper do not refer to the standard two-particle unitarity-cuts.}

Before closing, we motivate the most famous on-shell construction of massless scattering amplitudes: BCFW-recursion. In it, two null external momenta, $p_1^{\mu}$ and $p_2^{\mu}$, are deformed by a complex null-momentum, $z \times q^{\mu}$. The shift is such that (a) the shifted momenta $p_1(z) = p_1 + qz$ and $p_2(z) = p_2 - q z$ remain on-shell (possible, as momentum $q^{\mu}$ is complex), and (b) the total sum of external momenta remains zero.

As tree amplitudes are rational functions of their external kinematical invariants with, at most, simple poles, this deformation allows one to probe the analytic pole structure of the deformed amplitude, $A^{\rm tree}(z)$:
\begin{align}
A^{\rm tree}(z) \equiv A^{\rm tree}(p_1^{h_1}(z), p_2^{h_2}(z),...p_n^{h_n}) \, .
\end{align}
Kinematical poles in $A^{\rm tree}(z)$ are either un-shifted, or scale as $1/K^2 \rightarrow 1/(2 z (q \cdot K) + K^2)$, if $K$ includes only one of $\hat{p_1}$ or $\hat{p_2}$. Cauchy's theorem then gives a simple expression for the physical amplitude, $A^{\rm tree}(z = 0)$,
\begin{align}
A^{\rm tree}(z = 0) = \sum_{z_P} {\rm Res} \bigg\{ \frac{A_4(z)}{z} \bigg\}\bigg|_{z_P = -\frac{K^2}{2 q \cdot K}} +\big( {\rm Pole \, at} \, z\rightarrow \infty\big) \, .
\end{align}
Existence of such a BCFW-shift, in both Yang-Mills/QCD and in General Relativity, that dies off  at least as quickly as $1/z$ for large-$z$ can be elegantly shown through imposing complex factorization\cite{ST}, and allows the entire on-shell S-matrix to be built up from three-point amplitudes. Existence of valid BCFW-shifts were originally shown within local formulations of field theory~\cite{BCFW}\cite{BCFWinGR}\cite{SpinLorentz}. In section~\ref{niceshift} we develop a shift at four-points which is guaranteed to die off  for large-$z$ by simple dimensional analysis.

\section{Ruling out constructible theories by pole-counting}\label{PoleCount}

Pedestrian counting of poles, mandated by constructibility in four-point amplitudes, strongly constrains on-shell theories. The number of poles in an amplitude \emph{must} be less than or equal to the number of accessible, physical, factorization channels at four points. Tension arises, because the requisite number of poles in a four-point amplitude \emph{increases} with the highest-spin particle in the theory, while the number of possible factorization channels is \emph{bounded} from above by three, the number of Mandelstam variables.  

This tension explicitly rules out the following theories as inconsistent with constructibility, locality, and unitarity: (1) all relevant interactions (A = 0), save $\phi^3$ and an ``exotic'' Yukawa-like interaction, (2) all marginal interactions (A = 1) save those in YM, QCD, Yukawa theory, and scalar QED, and another ``exotic'' interaction between spin-$3/2$ particles and gluons, and (3) all first-order irrelevant interactions  (A = 2) save those in GR. Further consistency conditions later rule out those two unknown, pathological, relevant ($A = 0$) and marginal ($A = 1$) interactions. 

Further, incrementally more sophisticated pole-counting sharply constrains highly irrelevant ($A > 2$) higher-spin amplitudes. Specifically, save for two notable examples, they cannot consistently couple either to gravitational interactions or to more conventional Yang-Mills theories or ``gauge''-theories. This is the subject of section~\ref{UniqueLaws}. It is somewhat striking that simply counting poles in this way so powerfully constrains the palate of three-point amplitudes which may construct local and unitary S-matrices. The results of this pole-counting exercise are succinctly summarized in Fig.~\ref{HA1}.

\subsection{The basic consistency condition}

Explicitly we find that four-particle S-matrices constructed from primitive three-particle amplitudes are inconsistent with locality and unitarity if there are more than three poles in any given term in an amplitude. More specifically, the number of poles in the simplest amplitudes has to be \emph{at least} $N_p = 2H + 1 - A$. Thus, a theory is necessarily inconsistent if
\begin{align}
2H + 1 - A = N_p > 3 \, \Longleftrightarrow \, {\rm Number \, of \, poles } > {\rm cardinality \, of \, }\{s, t, u\} \, .
\label{Constraint1}
\end{align}
Recall that, in accordance with Eq.~\eqref{HA}, $A = | h_1 + h_2 + h_3| $ and $H = {\rm max}\{|h_1|,|h_2|,|h_3|\}$. 

We prove constraint~\eqref{Constraint1} below.\footnote{For expediency, we defer discussion of one technical point, proof of Eq.~\eqref{n4anatomy1}, to appendix~\ref{n4construct}.} First, we note there are $A$ total spinor-brackets in three-point amplitudes of the type in Eq.~\eqref{3ptA3}:
\begin{align}
A_3(1^{h_1},2^{h_2},3^{h_3}) = \kappa_A [12]^c [13]^b [23]^a \Rightarrow a+b+c = \sum_{i=1}^3 h_i = A > 0 \ .
\label{A3count2}
\end{align}
Thus, on a factorization channel of a four-point amplitude, $A_4$, \emph{constructed} from a given three-point amplitude multiplied by its parity conjugate amplitude, $A_3 \times \bar{A}_3$, there will be $A$ net holomorphic spinor-brackets and $A$ net anti-holomorphic spinor-brackets: $A \ \langle \ \rangle$s and $A \ [ \ ]$s.  Therefore, generically on such a factorization channel, the mass-squared dimension of the amplitude is:
\begin{align}
A_4 \rightarrow \frac{\kappa_A^2}{s_{\alpha \beta}} A_3 \times \bar{A}_3 \Rightarrow \bigg[\frac{A_4}{\kappa_A^2}\bigg] = (K^2)^{A-1} \ .
\label{A4count1}
\end{align}
By locality, an amplitude may only have $1/K^2$-type poles. Therefore the helicity information, captured by the non-zero little-group weight of the spinor-products, can only be present in an overall \emph{numerator} factor multiplying the amplitude. Four-point amplitudes thus naturally split into three parts: a numerator, $N$, which encodes helicities of the states, a denominator, $F(s,t,u)$, which encodes the pole-structure, and the coupling constants, $\kappa_A^2$, which encode the species-dependent characters of the interactions (discussed in section~\ref{niceshift}):
\begin{align}
A_4 = \kappa_A^2 \frac{N}{F(s,t,u)} \Rightarrow \bigg[ \frac{N}{F(s,t,u)} \bigg] = (K^2)^{A-1} \ ,
\label{A4anatomy1}
\end{align}
\begin{figure}
\begin{center}
\includegraphics{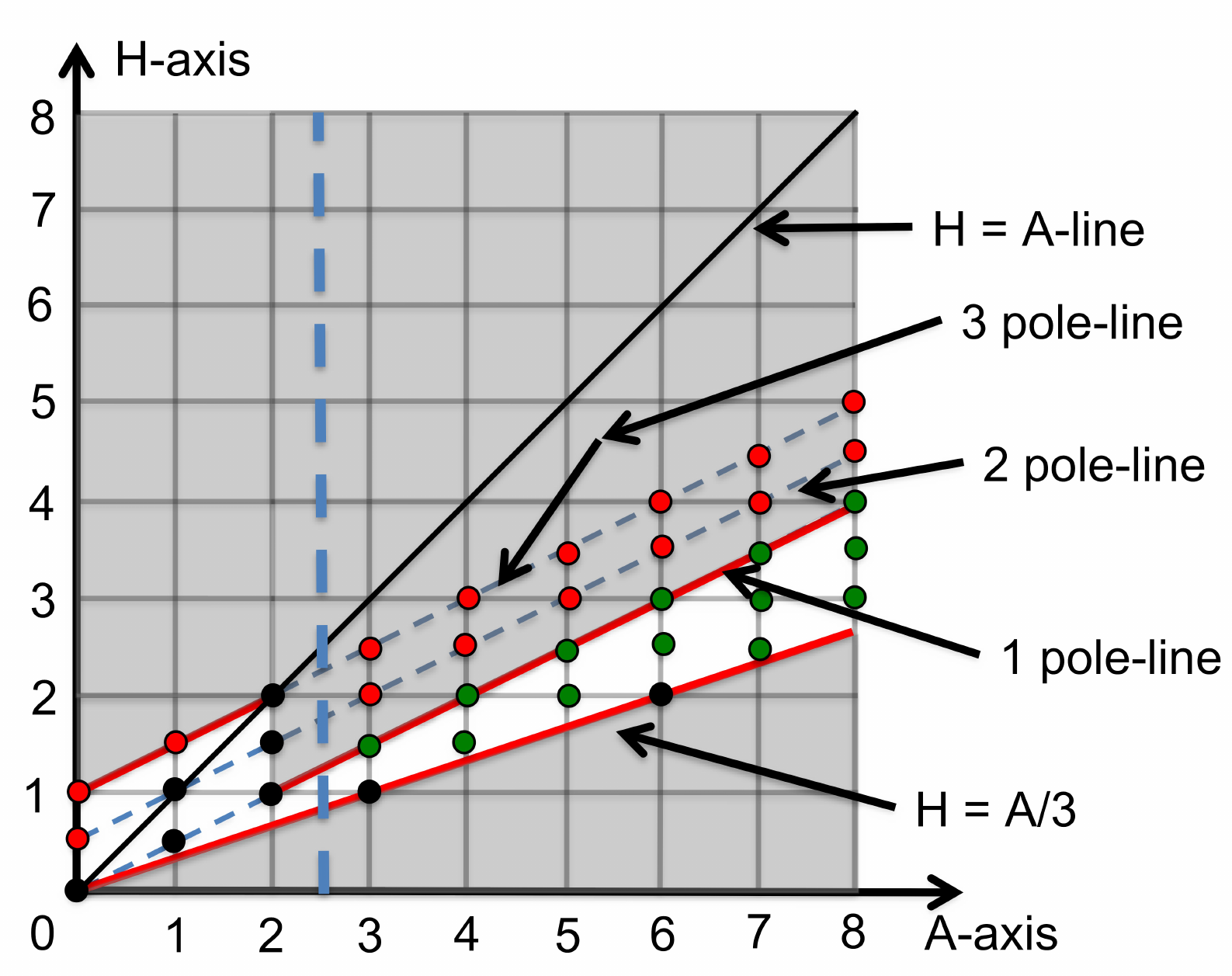}
\caption{Summary of pole-counting results. Recall $N_p = 2H + 1 - A$, where $A = |\sum_{i = 1}^3 h_i |$ and $H = {\rm max}\{ |h_i|\}$. (Color online.) In short: black-dots represent sets of three-point amplitudes that define self-consistent S-matrices that can couple to gravity; green-dots represent sets of three-point amplitudes which---save for two exceptions explicitly delineated in Eq.~\eqref{remnant}---define S-matrices that \emph{cannot} couple (in the sense defined in section~\ref{UniqueLaws}) to any S-matrix defined by the black-dots; red-dots represent sets of three-point amplitudes that cannot ever form consistent S-matrices. Straightforward application of constraint~\eqref{Constraint1}, in subsection~\ref{Big3}, rules out all $A_3$s with $(H,A)$-\emph{above} the $N_p = 3$-line. More careful pole-counting, in subsection~\ref{Np32} and appendix~\ref{Np2explicit}, rules out all interactions above the $N_p = 1$ line, save for those with $(H,A) = (1/2,0)$, $(1,1)$, $(3/2,2)$, and $(2,2)$. Further, in section~\ref{UniqueLaws}, a modified pole-counting rules out interaction between the $(H,A) = (2,2)$ gravity theory and any other theory with a spin-2 particle, save the unique $(H,A) = (2,6)$-theory. Similar results hold for gluon self-interactions: vectors present in any higher-spin amplitude with $A > 3$, save the unique $(H,A) = (1,3)$-theory, cannot couple to the vectors interacting via leading $(H,A) = (1,1)$ interactions. Section~\ref{weirdtheory} rules out the $(H,A) = (1/2,0)$-interaction. Amplitudes in the grey-shaded regions can never be consistent with locality and unitarity. Higher-spin, $A > 3$, amplitudes between the $H = A/2$ and $A = A/3$ lines may be consistent. However, they \emph{cannot} be coupled either to GR or YM, save for $(H,A) = (1,3)$ or $(2,6)$. In section~\ref{SUSY}, we show inclusion of leading-order interactions between massless spin-3/2 states, at $A = 2$, promotes gravity to supergravity. Supergravity cannot couple to even these two $A > 2$ interactions, as seen in appendix~\ref{SUGRAexclude}.}
\label{HA1}
\end{center}
\end{figure}where the last equality is inferred from Eq.~\eqref{A4count1}.  We prove in Appendix~\ref{n4construct}, that \emph{minimal} numerators $N$ which accomplish this goal are comprised of exactly $2H$ holomorphic and $2H$ anti-holomorphic spinor-brackets, \emph{none} of which can cancel against any pole in $F(s,t,u)$:
\begin{align}
N \sim \langle \, \rangle_{(1)} ... \langle \, \rangle_{(2H)} \,\,\, [ \, ]_{(1)} ... [ \, ]_{(2H)} \Rightarrow [N] = (K^2)^{2H} \ . 
\label{n4anatomy1}
\end{align}
Thus, by~\eqref{A4count1},~\eqref{A4anatomy1}, and~\eqref{n4anatomy1}, we see
\begin{align}
&\bigg[ \frac{A_4}{\kappa_A^2} \bigg] = \bigg[ \frac{N}{F(s,t,u)} \bigg] = (K^2)^{A-1} \, , \, {\rm and} \, [N] = (K^2)^{2H} \nonumber\\
& \qquad \qquad \Rightarrow [F(s,t,u)] = (K^2)^{2H+1-A}  \nonumber\\
& \qquad \qquad \Rightarrow N_p = 2H + 1 - A \, . 
\label{poleCount1}
\end{align}
Constraint~\eqref{Constraint1} naturally falls out from Eq.~\eqref{poleCount1}, after observing that there can be at most three legitimate, distinct, factorization channels in any four-point tree amplitude. This specific constraint, and others arising from pole-counting from minimal numerators, is extremely powerful. The catalogue of theories they together rule out are succinctly listed in Fig.~\ref{HA1}. We explore the consequences of this constraint below.

\subsection{Relevant, marginal, and (first-order) irrelevant theories ($A \leq 2$): constraints}\label{Big3}

To begin with, note that constraint~\eqref{Constraint1} immediately rules out all theories with $N_p > 3$. Beginning with relevant, $A = 0$, interactions, we see that $N_p = 2H + 1 \leq 3 \Rightarrow H \leq 1$. Already this rules out relevant interactions between massless spin-3/2 and spin-2 states. 

Next, argument by contradiction rules out relevant amplitudes involving massless vectors, i.e. the $(H,A) = (1,0)$-theories wh. Consider such a relevant amplitude, for example $A_3(+1,-1/2,-1/2)$. It constructs a putative four-point amplitude with external vectors,
\eq
A_4\left(-1,-\frac{1}{2},\frac{1}{2},1\right) \label{shitty1} \, .
\eqe
This amplitude must have $2 + 1 - 0 = 3$ poles, each of which must have an interpretation as a \emph{valid} factorization channel of the amplitude. So it must have valid $s$-, $t$-, and $u$-factorization channels, with \emph{relevant} ($A = 0$) three-point amplitudes on either side. However, on the $s \rightarrow 0$ pole, $A_4$ factorizes as,
\eq
A_4 \left(-1,-\frac{1}{2},\frac{1}{2},1\right) \bigg|_{s \rightarrow 0} = \frac{1}{s} 
\bar{A}_3 \left(-1,-\frac{1}{2}, h\right) 
A_3 \left(1, \frac{1}{2}, -h\right)  \label{shitty2} \, .
\eqe
where $h$ must be 3/2 to make the interaction relevant. Thus, to be consistent with locality and unitarity, relevant vector couplings must also include spin-3/2 particles. But, as mentioned above, including these particles in the spectrum, and then taking them as external state invariably leads to too many poles. An identical argument shows that the remaining relevant vertex $A_3(+1,-1,0)$ requires spin 2 particles, again leading to an inconsistency. Thus all $(H,A) = (1,0)$ interactions are also ruled out. Thus the only admissible relevant three-point amplitudes are
\eq
A_3(0,0,0) \,\, , \,\, {\rm and} \,\, A_3\left(0,\frac{1}{2},-\frac{1}{2}\right) \, .
\eqe
The first amplitude is the familiar one from $\phi^3$-theory. We rule out the second amplitude in section~\ref{weirdtheory}.

Further, we see that marginal interactions cannot contain particles with helicities larger than $3/2$. Directly, requiring $2H + 1 - A \leq 3$ for $A = 1$ forces $H \leq 3/2$. $(H,1)$-type three-point amplitudes cannot build S-matrices consistent with locality and unitarity for $H > 3/2$.

We rule out \emph{marginal} $(H,A) = (3/2,1)$ amplitudes, i.e. marginal coupling to massless spin-3/2 states, using the same logic as above. This time, marginal amplitudes with external 3/2 particles require \emph{all} three poles. Two factorization channels have consistent interpretations within the theory; however, the ``third'' channel does not. It necessitates exchange of a spin-2 state between the three-point amplitudes. But this violates constraint~\eqref{Constraint1}: marginal amplitudes with spin-2 states lead to amplitudes with four kinematic poles. Thus, the only admissible marginal three-point amplitudes are,
\begin{align}
A_3\left(1,1,-1\right) \, , \,\, 
A_3\left(1,\frac{1}{2},-\frac{1}{2}\right) \, , \,\, 
A_3\left(1,0,0\right) \, , \,\, {\rm and} \,\, 
A_3\left(0,\frac{1}{2},\frac{1}{2}\right) \, , \label{SYMint}
\end{align} 
and their conjugate three-point amplitudes. We refer to this set of three-point amplitudes, loosely, as ``the ${\cal N} = 4$ SYM interactions."

Finally, constraint~\eqref{Constraint1} rules out leading-order gravitational coupling to particles of spin-$H > 2$. Such three-point amplitudes, of the form $A_3(H, -H, \pm2)$, have $A = 2$ and $H > 2$, and yield four-point amplitudes with $2H -1 > 3$ poles; this cannot be both unitary and local for $H > 2$. Admissible $A = 2$ amplitudes are restricted to:
\begin{align}
A_3\left(2,2,-2\right) \, , \,\, 
A_3\left(2,\frac{3}{2},-\frac{3}{2}\right) \, , \,\, 
A_3\left(2,1,-1\right) \, , \,\, 
A_3\left(2,\frac{1}{2},-\frac{1}{2}\right) \, , \,\, 
A_3\left(2,\frac{1}{2},-\frac{1}{2}\right) \, , \label{GRint} \\
A_3\left(\frac{3}{2},\frac{3}{2},-1\right) \, , \,\, 
A_3\left(\frac{3}{2},1,-\frac{1}{2}\right) \, , \,\, 
A_3\left(\frac{3}{2},\frac{1}{2},0\right) \, , \,\,
A_3\left(1,\frac{1}{2},\frac{1}{2}\right)\, , \,\, {\rm and} \,\, 
A_3\left(1,1,0\right) \, , \label{SUGRAint}
\end{align} 
and their conjugate three-point amplitudes. We refer to the amplitudes in~\eqref{GRint} as ``gravitational interactions.'' More generally, we refer to this full set of three-point amplitudes, loosely, as ``the ${\cal N} = 8$ SUGRA interactions."

\subsection{Killing $N_p = 3$ and $N_p = 2$ theories for $A \geq 3$}\label{Np32}

It is relatively simple to show that any theory constructed from $A_3$s with $N_p = 3$ poles, beyond $A = 2$, cannot be consistent with unitarity and locality. To begin, we note that
\begin{align}
\big\{N_p = 3 \Longleftrightarrow 2H + 1 - A = 3 \big\} \Rightarrow H = A/2 + 1 \label{3kill1} \, .
\end{align}
We label the helicities in the three-point amplitudes with $N_p = 3$, as $A_3(H,g,f)$. Without loss of generality, we order them as $f \leq g \leq H = A/2 + 1$. As $A > 2$, then $g + f$ must be positive: at a minimum $g > 0$. 

Now construct the four-point amplitude $A_4(H, -H, f, -f)$ from this three-point amplitude and its parity-conjugate. By assumption, this amplitude must have three poles, each of which must have an interpretation as a legitimate factorization channel within the theory constructed from $A_3(A/2 + 1, g, f)$ (or some mild extension of the theory/spectrum).

However, in order for the $t$-channel pole in the amplitude $A_4(H, - H, g, -g)$ to have a viable interpretation as a factorization channel, it requires a state with spin \emph{greater} than $A/2+1 = H$. Specifically, on this $t$-pole,
\eq
A_4(H, - H, g, -g)\bigg|_{t \rightarrow 0} = \frac{1}{K_{14}^2} A_3\bigg(\frac{A + 2}{2}, -g, \frac{A+2}{2}+g\bigg) \bar{A}_3\bigg(-\frac{A + 2}{2}, g, -\frac{A+2}{2}-g\bigg)  
\label{3kill2}
\eqe
By assumption, $g > 0$: the intermediate state must have helicity $\tilde{H} = A/2 + 1 + g$. Clearly this new state has helicity larger than $H = A/2 + 1$. A priori, there is no problem: new particles mandated by consistency conditions may be included into the spectrum of a theory without necessarily introducing inconsistencies. However, if \emph{these} particles of spin $\tilde{H} > H = A/2 + 1$ are put as external states of the new three-point amplitudes in the modified theory, then these new four-point amplitudes will necessarily have $3 + 2g > 3$ poles, in violation of constraint~\eqref{Constraint1}.

Hence all theories constructed from $A_3$s with $N_p \geq 3$ and $A>2$ are inconsistent. Similar arguments show that theories with $N_p = 2$ cannot be consistent for $A>2$; they are however slightly more detailed, and involve several specific cases at low-$A$ values. Proof of this extended claim is relegated to Appendix~\ref{Np2explicit}.

\section{There is no GR (YM) but the true GR (YM)}\label{UniqueLaws}

In this section, we investigate further constraints imposed by coupling $A\ge3$ theories to GR (YM) interactions. This is done by considering four-point amplitudes which factorize as $A_4 \rightarrow A_{GR} \times A_3$ and $A_4 \rightarrow A_{YM} \times A_3$, where $A_3$ is the vertex of some other theory. Note however that the arguments in this section apply only to three-point interactions which contain either a spin-2 or a spin-1 state. Other higher spin theories are not constrained in any way by this reasoning.

First, we find that all higher-spin theories are inconsistent if coupled to gravity. This is in addition to the previous section, where spin $s>2$ theories with $A>2$ were allowed if $N_p\le1$. Further, we show that massless spin-2 states participating in $A > 2$ three-point amplitudes must be identified with \emph{the} graviton which appears in the usual $A=2$ $A_3(+2,-2,\pm2)$ three-point amplitudes defining the S-matrix of General Relativity. Pure pole-counting shows that \emph{no} massless spin-2 state in \emph{any} three-point amplitude with $A > 2$ can couple to GR, unless they are within the unique $(H, A) = (2, 6)$ three-point amplitudes, $A_3(2, 2, 2)$ and its complex conjugate. Similar results hold for gluons.\footnote{We further show, in appendix~\ref{UniqueSUGRA} that theories with spin-3/2 states are also unique in a similar manner.}

To rule out higher-spin theories interacting with gravity, we show that amplitudes with factorization channels of the type,
\begin{align}
A_4(1^{+2},2^{-2},3^{-H},4^{-h}) \rightarrow \frac{1}{K^2} A_3(2,-2,+2) \times A_3(-2,-H,-h) \label{GRfact1}
\end{align}
cannot be consistent with unitarity and locality, unless $|H| \leq 2$ and $|h| \leq 2$. 

It is relatively easy to see this, especially in light of the constraints from sections~\ref{Big3} and~\ref{Np32}, which fix $H \leq A/2$ for $A \geq 3$. Note that, in order to even couple to GR's defining three-graviton amplitude, the three-point amplitude in question must have a spin-2 state. These two conditions admit only three possible three-point amplitudes, for a given $A$:
\begin{align}
A_3(A/2-1,A/2-1,2)        &\Rightarrow A_4(1^{+2},2^{-2},3^{-(A/2-1)},4^{-(A/2-1)} )       \, , \nonumber\\ 
A_3(A/2-1/2,A/2-3/2,2) &\Rightarrow A_4(1^{+2},2^{-2},3^{-(A/2-1/2)},4^{-(A/2-3/2)} ) \, , \,\, {\rm and} \label{GRa4s} \\ 
A_3(A/2,A/2-2,2)           &\Rightarrow A_4(1^{+2},2^{-2},3^{-(A/2)},4^{-(A/2-2)} )                     \, . \nonumber
\end{align}
The minimal numerator which encodes the spins of the extenal states in, for instance, the first amplitude, must be,
\begin{align}
N \sim [1| P | 2 \rangle^4 \left(\langle 34 \rangle^2\right)^{(A/2 - 1)} \Rightarrow [N] = (K^2)^{3+A/2} \, . \label{GRnum1}
\end{align}
However, by power-counting, the kinematic-dependent part of the amplitude must have mass-dimension,
\begin{align}
\bigg[\frac{N}{f(s,t,u)} \bigg] = \bigg[ \frac{1}{K^2} A_{\rm Left}^{(GR)} A_{\rm Right}^{(A)} \bigg] = \frac{(K^2)^{2/2} \, (K^2)^{A/2}}{(K^2)}  = (K^2)^{A/2}\, , \label{GRk2}
\end{align}
and thus the denominator, $f(s,t,u)$, must have mass-dimension,
\begin{align}
[f(s,t,u)] = (K^2)^3 \Rightarrow f(s,t,u) = s \, t \, u \, ! \label{GRpoleCount}
\end{align}
Casual inspection shows us that the ``third'' factorization channel, to be sensible, requires an intermediary with spin $A/2-1$ to couple directly via the leading $A = 2$ gravitational interactions. This, and similar analysis for the other two classes of three-point amplitudes in Eq.~\eqref{GRa4s}, proves that the spin-2 particle associated with the graviton in the leading-order, $(H,A) = (2,2)$, gravitational interactions can only participate in three higher-derivative three-point amplitudes, namely, 
\eq
A_3\left( +2,+1,+1\right) \,,\, A_3\left(+2,+\frac{3}{2},+\frac{3}{2}\right) \,,\, A_3\left(+2,+2,+2\right) \label{remnant}
\eqe
In the special case of the three-point amplitude $A_3(+2,+2,+2)$, the third channel simply necessitates an intermediate spin-2 state, the ``graviton.'' Thus GR can couple to itself, or amplitudes derived from $R^a_{\,\,\,b} R^b_{\,\,\,c} R^c_{\,\,\,a}$, its closely related higher-derivative cousin~\cite{SoftGR}\cite{Weinberg1}.\footnote{The minimal numerators for the other candidate amplitudes in this theory, Eq.~\eqref{GRa4s}, have the same number of spinor-brackets in their numerator as that in Eq.~\eqref{GRnum1}; thus have the same mass-dimensions. Therefore all amplitudes must identical number of poles, and as in Eq.~\eqref{GRpoleCount}, they have $f(s,t,u) = s t u$.}

Second, gluons. Specifically, we show that gluons, i.e. the massless spin-1 particles which couple to each-other at \emph{leading} order via the $H = A = 1$ three-point amplitudes, can \emph{not} consistently couple to any spin $s > 1$ within $A \geq 3$ amplitudes. This means that any constructible amplitude with factorization channels of the type,
\begin{align}
A_4(1^{+1},2^{-1},3^{-H},4^{-h}) \rightarrow \frac{1}{K^2} A_3(1,-1,+1) \times A_3(-1,-H,-h) \label{YMfact1}
\end{align}
cannot be consistent with unitarity and locality, unless $|H| \leq 1$ and $|h| \leq 1$. 

Again, in light of the constraints from sections~\ref{Big3} and~\ref{Np32}, which fix $H \leq A/2$ for $A \geq 3$, it is relatively easy to see this. To even possibly couple to this three-gluon amplitude, the three-point amplitude in question must have a spin-1 state. These two conditions allow only two possible three-point amplitudes, for a given $A$:
\begin{align}
A_3(A/2-1/2,A/2-1/2,1) &\Rightarrow A_4(1^{+1},2^{-1},3^{-(A/2-1/2)},4^{-(A/2-1/2)} ) \, , \,\, {\rm and} \nonumber\\ 
A_3(A/2,A/2-1,1)           &\Rightarrow A_4(1^{+1},2^{-1},3^{-(A/2)},4^{-(A/2-1)} )                     \, . \label{YMa4s}
\end{align}
The minimal numerator which encodes the spins of the extenal states in, for instance, the first amplitude, must be,
\begin{align}
N \sim [1| P | 2 \rangle^2 (\langle 34 \rangle^2)^{(A/2 - 1/2)} \Rightarrow [N] = (K^2)^{A/2+3/2} \, . \label{YMnum1}
\end{align}
However, by power-counting, the kinematic-dependent part of the amplitude must have mass-dimension,
\begin{align}
\bigg[\frac{N}{f(s,t,u)} \bigg] = \bigg[ \frac{1}{K^2} A_{\rm Left}^{(YM)} A_{\rm Right}^{(A)} \bigg] = \frac{(K^2)^{1/2} \, (K^2)^{A/2}}{(K^2)} = (K^2)^{A/2-1/2} \, , \label{YMk2}
\end{align}
and thus the denominator, $f(s,t,u)$ must have mass-dimension two:
\begin{align}
[f(s,t,u)] = (K^2)^2 \Rightarrow 1/f(s,t,u) \, {\rm must \, have \, at \, least \,two \, poles}  . \label{YMpoleCount}
\end{align}
Again, casual inspection shows that, while the one pole---that in Eq.~\eqref{YMfact1}---indeed has a legitimate interpretation as a factorization channel within this theory, the ``second'' channel generically does not: it requires the gluon to \emph{marginally} couple to spin $A/2 - 1/2 \geq 1$ states. As seen in section~\ref{Big3}, this cannot happen---unless $A/2 - 1/2 = 1 \Leftrightarrow A = 3$. 

For the second amplitude in  Eq.~\eqref{YMa4s} the argument is a bit more subtle when $A=3$. In this case, the $u$-channel is prohibited, but the $t$-channel is valid:
\begin{align}
A_4(1^{+1},2^{-1},3^{-3/2},4^{-1/2}) \rightarrow \frac{1}{K^2}A_3(1,-1/2,1/2)\times A_3(-1/2,-3/2,-1) \label{weirdguy}
\end{align}
This interaction is ruled out through slightly more detailed arguments, involving the structure of the vector self-coupling constant in $A_3(1,-1,\pm1)$---discussed in section~\ref{niceshift}. We pause to briefly describe how this is done, but will not revisit this particular, $A_3(1,-1/2,1/2)$ interaction further (it is just a simple vector-fermion QED or QCD interaction). Simply, we note that $A_3(1,-1,\pm1) \propto f_{abc}$, the structure-constant for a simple and compact Lie-Algebra; see Eq.~\eqref{YMcouple}. From here, it suffices to note that either by choosing the external vectors to be photons, or gluons of the same color, this amplitude vanishes, and then so does the original $s$-channel. Nothing is affected in Eq.~\eqref{weirdguy} and so Eq.~\eqref{YMpoleCount} cannot be fulfilled, implying that the coupling constant of $A_3(1,1/2,3/2)$ must vanish.

Thus at four-points YM can only couple to itself, gravity via the $A_3(\pm 2, 1, -1)$ three-point amplitude, or amplitudes derived from $F^a_{\,\,\,b} F^b_{\,\,\,c} F^c_{\,\,\,a}$, its closely related higher-derivative cousin.

\section{Behavior near poles, and a possible shift}\label{niceshift}

In this section we explain a new shift which is guaranteed to vanish at infinity. Using this shift, we re-derive classic results, such as (a) decoupling of multiple species of massless spin-2 particles~\cite{WW}, (b) spin-2 particles coupling to all particles (with $|H| \leq 2$, of course!) with identical strength, $\kappa = 1/M_{pl}$, (c) Lie Algebraic structure-constants for massless spin-1 self-interactions, and (d) arbitrary representations of Lie Algebra for interactions between massless vectors and massless particles of helicity $|H| \leq 1/2$.

Note that in section~\ref{PoleCount}, we proved that a four point-amplitude, constructed from a given three-point amplitude and its parity conjugate, $A_3^{(H,A)}$ and $\bar{A}_3^{(H,A)}$, takes the generic form,
\begin{align}
A_4 \sim \frac{\big( \, \langle \, \, \, \rangle [ \, \, \, ] \, \big)^{2H}}{(K^2)^{2H-A+1}} \, . 
\end{align}
Consequently in the vicinity of, say, the $s$-pole, the four-point amplitude behaves as,
\begin{align}
{\rm lim}_{s \rightarrow 0}A_4 = \frac{1}{s} \frac{ N }{t^{2H-A}} \, , \,\, {\rm where} \,\, N \sim ( \, \langle \, \, \, \rangle [ \, \, \, ] \, \big)^{2H} \, . 
\end{align}
We exploit this scaling to identify a useful shift that allows us to analyze constraints on the coupling-constants, the ``$g_{abc}$''-factor in three-point amplitudes [see Eq.~\eqref{3ptA3}].  Complex deformation of the Mandelstam invariants, which we justify in appendix~\ref{Illustration}, for arbitrary $\tilde{s}$ and $\tilde{t}$,
\begin{align}
(s, t, u) \rightarrow (s + z \tilde{s}, t + z \tilde{t}, u + z\tilde{u}) \ ,
\label{defMand1}
\end{align}
grants access to the poles of $A_4(s,t,u)$ without deforming the numerator. Partitioning,
\begin{align}
A_4 (z = 0) \sim \kappa_A^2 \frac{ N}{f(s,t,u)} \rightarrow A_4 (z) \sim \kappa_A^2 \frac{ N }{f(s(z),t(z),u(z))} \, , 
\end{align}
accesses the poles in each term, while leaving the helicity-dependent numerator \emph{un-shifted}. Basic power-counting implies that four-point amplitudes, constructed from three-point amplitudes of the type $A_3^{(H,A)} \times \bar{A}_3^{(H,A)}$, die off as $z \rightarrow \infty$ for $2H - A = 1, 2$ under this shift. Thus, four-point amplitudes are uniquely fixed by their finite-$z$ residues under this deformation:
\begin{align}
A_4(z=0) = \sum_{z_P} {\rm Res}\bigg(\frac{A_4(z)}{z}\bigg) \ .
\label{defMand2}
\end{align}
Straightforward calculation of the residues on the $s$-, $t$-, and $u$-poles yields,
\eq
A(1_a,2_b,3_c,4_d)=
\bigg\{
\frac{(\tilde{s})^{2H-A}}{s }g^{abi}g^{icd}+
\frac{(\tilde{t} )^{2H-A}}{t  }g^{adi}g^{ibc}+
\frac{(\tilde{u})^{2H-A}}{u}g^{aci}g^{ibd}
\bigg\} \frac{\rm Num}{(\tilde{s} t- \tilde{t} s)^{2H-A}} \, . \label{Spurious}
\eqe
Notably, this closed-form expression for the amplitude contains a non-local, spurious, pole which depends explicitly on the shift parameters, $\tilde{s}$ and $\tilde{t}$ (note: $\tilde{u} = -\tilde{s}-\tilde{t}$). Requiring these spurious parameters to cancel out of the final expression in theories, of self-interacting spin-1 particles, forces the Lie Algebraic structure of Yang-Mills ~\cite{BC}\cite{ST}. Similarly, for theories of interacting spin-2 particles, we recover the decoupling of multiple species of massless spin-2 particles\cite{WW}, and the equal coupling of all spin $|H| < 2$ particles to a spin-2 state~\cite{SoftGR}\cite{BC}\cite{ST}.

\subsection{Constraints on vector coupling $(A = 1)$}

Here we derive consistency conditions on Eq.~\eqref{Spurious} for scattering amplitudes with external vectors, interacting with matter via leading-order, $A = 1$, couplings; $2H - A = 1$. Now, if the amplitude is invariant under changes of $\tilde{s} \rightarrow \tilde{S}$, then it necessarily follows that the same holds for re-definitions $\tilde{t} \rightarrow \tilde{T}$, and thus that unphysical pole cancels out of the amplitude.

Therefore, if $\frac{\partial A}{\partial \tilde{s}}=0$, then it indeed follows that the amplitude is invariant under redefinitions of the shift parameter, $\tilde{s}$, and the unphysical pole has trivial residue. Beginning with the all-gluon amplitude, where the three-point amplitudes are $A_3(1_a^{+1},2_b^{-1},3_c^{\pm 1}) \propto f_{abc}$, we see that the derivative is, 
\begin{align} 
\frac{\partial A_4}{\partial \tilde{s}}\bigg|_{(H,A) = (1,1)} \propto f^{abi}f^{icd} + f^{aci}f^{ibd} + f^{adi}f^{ibc}  \, . \label{YMcouple}
\end{align} 
Requiring this to vanish is equivalent to imposing the Jacobi identity on these $f_{abc}$s. Thus, requiring the amplitude to be physical forces the gluon self-interaction to be given by the adjoint representation of a Lie group~\cite{BC}\cite{ST}\cite{PITP}.

Next, considering four-point amplitudes with two external gluons and two external fermions or scalars, we are forced to introduce a new type of coupling: $A_3(1_a^{\pm1},2_b^{+h},3_c^{-h}) \propto (T_a)_{bc}$. Concretely, we wish to understand the invariance of $A_4(1_a^{+1},2_b^{-1},3_c^{-h},4_d^{+h})$, constructed from the shift~\eqref{defMand1}, under redefinitions $\tilde{s} \rightarrow \tilde{S}$. 

Factorization channels on the $t$- and $u$-poles are given by the products of two $A_3$s with one gluon and two spin-$h$ particles, and thus are proportional to $(T_a)_{ci} (T_b)_{di}$ and $(T_a)_{di} (T_b)_{ci}$, respectively--while the $s$-channel is proportional to $f_{abi} (T_i)_{cd}$. So, $\frac{\partial A}{\partial \tilde{s}}$ is proportional to,
\begin{align} 
(T_a)_{ci} (T_b)_{id} - (T_a)_{di} (T_b)_{ic} + f_{abi} (T_i)_{cd}  \, . \label{QCDcouple}
\end{align} 
This is nothing other than the definition of the commutator of two matrices, $T_{a}$ and $T_{b}$ in an arbitrary representation of the Lie group ``defined'' by the gluons in Eq.~\eqref{YMcouple}~\cite{BC}\cite{ST}\cite{PITP}.

\subsection{Graviton coupling}

Four-point amplitudes with two external gravitons have $2H - A = 2$, and so Mandelstam deformation~\eqref{defMand1} yields,
\eq
A(1_a^{-2},2_b^{-h},3_c^{+2},4_d^{+h})=
\bigg\{
\frac{\tilde{s}^{2}}{s}  \kappa_{h}^{abi}     \kappa_{h}^{icd}   +
\frac{\tilde{t}^{2}}{t}    \kappa_{h}^{adi}     \kappa_{h}^{ibc} +
\frac{\tilde{u}^{2}}{u} \kappa_{h=2}^{aci} \kappa_{h}^{ibd}
\bigg\} \frac{(\langle 1 2 \rangle [34])^{4-2h} \langle 1 | 2-3 | 4 ]^{2h}}{(\tilde{s} t- \tilde{t} s)^{2}} \, , \label{GRcouple1}
\eqe
where $\kappa_h^{abc}$ is the coupling constant in $A_3(1_a^{\pm 2},2_b^{-h},3_c^{+h})$. Demanding this amplitude be independent of redefinitions of $\tilde{s} \rightarrow \tilde{S}$, again reduces to the constraint that the partial derivative of Eq.~\eqref{GRcouple1} must vanish. Evaluating the derivative, we see,
\eq
\frac{\partial A_4}{\partial \tilde{s}}\bigg|_{(H,A) = (2,2)} \propto 
\tilde{s}\left(\kappa_{h}^{abi}\kappa_h^{icd} - \kappa_{h=2}^{adi}\kappa_h^{ibc}\right) + 
\tilde{t} \left(\kappa_{h}^{aci}\kappa_h^{ibd} - \kappa_{h=2}^{adi}\kappa_h^{ibc}\right)  \, , \label{GRcouple2}
\eqe
which vanishes only if,
\eq
\kappa_{h}^{abi}\kappa_{h}^{icd} = 
\kappa_{h=2}^{adi}\kappa_{h}^{ibc} \,\, , \, {\rm and} \,\, 
\kappa_{h}^{aci}\kappa_{h}^{ibd} =
\kappa_{h=2}^{adi}\kappa_{h}^{ibc} \, . \label{GRcouple}
\eqe
As noted in~\cite{BC}, for $h = 2$, this implies that the $\kappa_h^{abc}$s are a representation of a commutative, associative algebra. Such algebras can be reduced to self-interacting theories which decouple from each other. In other words, multiple gravitons, i.e. species of massless spin-2 particles interacting via the leading-order $(H,A) = (2,2)$ three-point amplitudes, necessarily decouple from each-other. This is the perturbative casting of the Weinberg-Witten theorem~\cite{WW}. As the multiple graviton species decouple, we refer to the \emph{diagonal} graviton self-interaction coupling as, simply, $\kappa$.

Diagonal gravitational self-coupling powerfully restricts the class of solutions to Eq.~\eqref{GRcouple} for $h < 2$. Directly, it implies that any individual graviton can only couple to a particle-antiparticle pair. In other words, $\kappa_{h}^{g ab} = 0$, for different particle flavors {\it a} and {\it b} on the spin-$\pm h$ lines. Similar to the purely gravitational case, we write simply $\kappa_h^{g aa} = \kappa_h$. Furthermore, to solve Eq.~\eqref{GRcouple} for $h \neq 2$ then it also must hold that $\kappa_h = \kappa_{h = 2} = \kappa$. In other words, the graviton self-coupling constant $\kappa$ is a simple constant; all particles which interact with a given unique graviton do so diagonally and with identical strengths. Thus multiple graviton species decouple into disparate sectors, and, within a given sector, gravitons couple to all massless states with identical strength, $\kappa$---the perturbative version of the equivalence principle~\cite{SoftGR}\cite{BC}\cite{ST}.

\subsection{Killing the relevant $A_3\big(0,\frac{1}{2},-\frac{1}{2}\big)$-theory}\label{weirdtheory}

This shift neatly kills the S-matrix constructed from the three-point amplitudes $A_3(\frac{1}{2},-\frac{1}{2},0)$. Just as before, we will see that in order for $A_4(0,0,\frac{1}{2},-\frac{1}{2})$ to be constructible (via complex Mandelstam - deformations) and consistent, the coupling constant in the theory must vanish.

As in YM/QCD, in this theory $2H - A = 1$. Invariance of $A_4(0,0,\frac{1}{2},-\frac{1}{2})$ under deformation redefinitions $\tilde{s} \rightarrow \tilde{S}$ again boils down to a constraint akin to Eq.~\eqref{YMcouple}---with one exception. Namely, there are only two possible factorization channels in this theory and not three: any putative $s$-channel pole would require a $\phi^3$ interaction, not present in this minimal theory. And so invariance under redefinitions $\tilde{s} \rightarrow \tilde{S}$ reduces to,
\begin{align}
\frac{\partial A_4}{\partial \tilde{s}}\bigg|_{(H,A) = (\frac{1}{2},0)} = f^{acp}f^{bdp} + f^{adp}f^{bcp} \, . \label{DEATH2}
\end{align}
The only solution to this constraint is for $f^{acp}f^{bdp} = 0 = f^{adp}f^{bcp}$, i.e. for the coupling constant to be trivially zero.\footnote{One may wonder why such an argument does not also rule out conventional well-known theories, such as spinor-QED or GR coupled to spin-1/2 fermions, as inconsistent. The resolution to this question is subtle, but boils down to the fact that amplitudes involving fermions in these $A > 0$ theories have extra, \emph{antisymmetric} spinor-brackets in their numerators. These extra spinor-brackets introduce a relative-sign between the two terms, and in effect modify the condition~\eqref{DEATH2} from $\left\{ff + ff = 0 \Rightarrow f = 0\right\}$ to $ff - ff = 0$, which is trivially satisfied.}

\section{Interacting spin-$\frac{3}{2}$ states, GR, and supersymmetry}\label{SUSY}

Supersymmetry automatically arises as a consistency condition on four-point amplitudes built from leading-order three-point amplitudes involving spin-3/2 states. In a sense, this should be more-or-less obvious from inspection of the leading-order spin-3/2 amplitudes in Eqs.~\eqref{GRint}, and~\eqref{SUGRAint}. For convenience, they are,
\begin{align}
A_3\left(\frac{3}{2},\frac{1}{2},0\right) \, , \,\, 
A_3\left(\frac{3}{2},1,-\frac{1}{2}\right) \, , \,\, 
A_3\left(\frac{3}{2},\frac{3}{2},-1\right) \, , \,\, {\rm and} \,\,
A_3\left(\frac{3}{2},2,-\frac{3}{2}\right)  \, . \label{GravitinoInt}
\end{align}
Clearly, every non-gravitational $A = 2$ amplitude with a spin-3/2 state involves one boson and one fermion, with helicity (magnitudes) that differ by exactly a half-unit. This should be unsurprising, as $A - 3/2 = 1/2$. Nonetheless, we should expect supersymmetry to be an emergent phenomena: throughout the previous examples, mandating a unitary interpretation of a factorization channel within novel four-point amplitudes in a theory forced introduction of new states with new helicities into the spectrum/theory. In a sense, the novelty of $A = 2$ amplitudes with external spin-3/2 states is that these new helicities do \emph{not} lead to violations of locality and unitarity. 

In amplitudes with external spin-3/2 states (and no external gravitons), each term in the amplitude must have $2H + 1 - A \rightarrow 3 + 1 - 2 = 2$ poles. Generically, one of these two poles will mandate inclusion of states with \emph{new} helicities into the spectrum of the theory. Fundamentally, we see that the \emph{minimal} $A = 2$ theory with a single species of spin-3/2 state is given by the two three-point amplitudes (and and their parity-conjugates):
\eq
A_3\left(\frac{3}{2},2,-\frac{3}{2}\right) \, , \,\, {\rm and} \,\, A_3(2,2,-2) \, .
\eqe
These interactions \emph{define} pure ${\cal N} = 1$ SUGRA, and are indicative of all other theories which contain massless spin-3/2 states (at leading order).  All non-minimal extensions of any theory containing spin-3/2 states necessarily contain the graviton. As we will make precise below, supersymmetry necessitates gravitational interactions---supersymmetry requires the graviton.

\begin{figure}
\begin{center}
\includegraphics{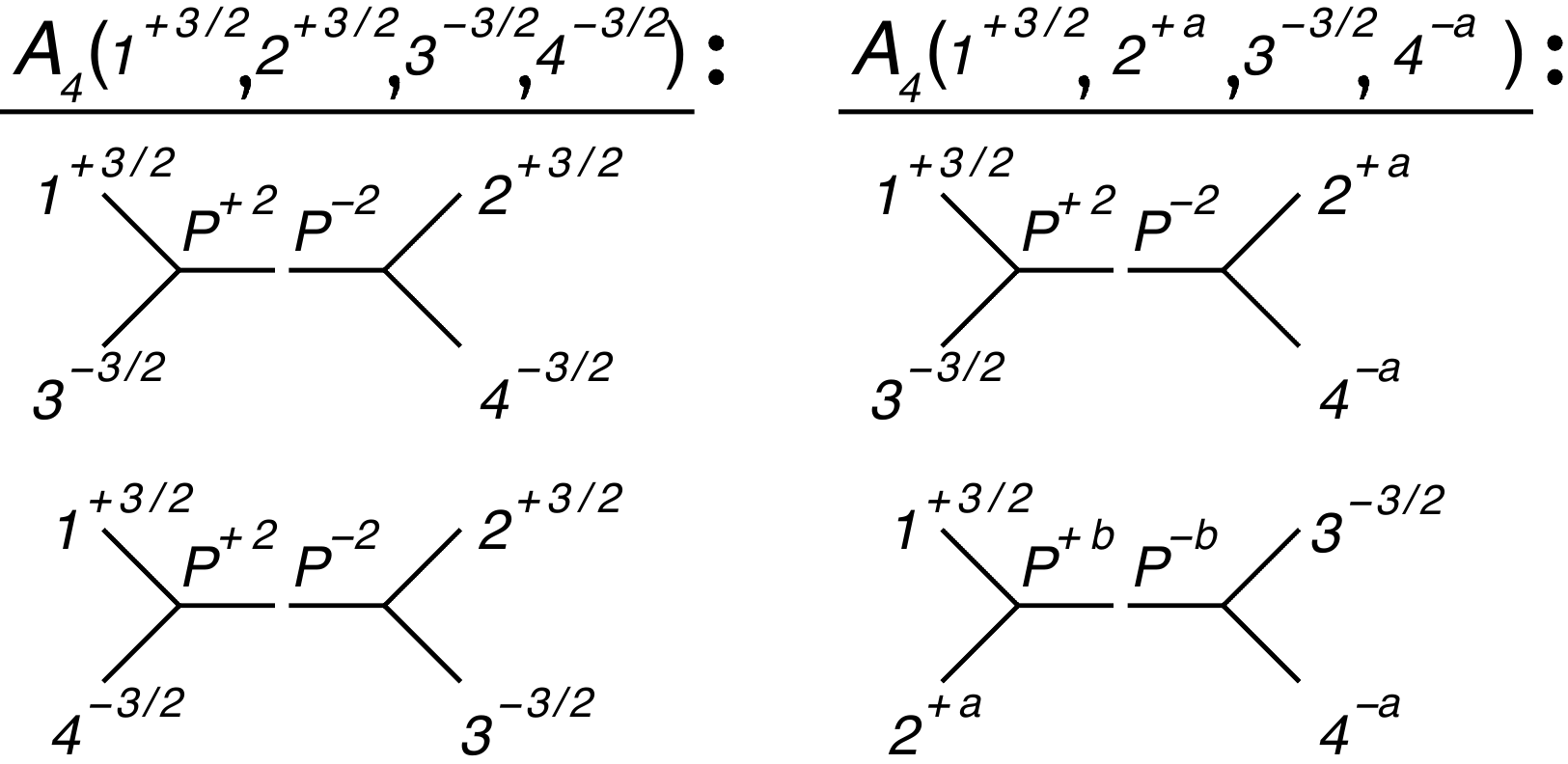}
\caption{Factorization necessitates gravitation in theories with massless spin-3/2 states. Specifically, figure (a)  represents the two factorization channels in the minimal four-point amplitude, $A_{4}(1^{+\frac{3}{2}},2^{+\frac{3}{2}},3^{-\frac{3}{2}},4^{-\frac{3}{2}})$ in an S-matrix involving massless spin-3/2 states. Further, figure (b) shows the two factorization channels present in the amplitude $A_4(3/2,-3/2, +a,-a)$.}
\label{SUGRAfact}
\end{center}
\end{figure}

Minimally, consider a four-particle amplitude which ties together four spin-3/2 states, two with helicity $h=+3/2$, and two with helicity $h=-3/2$, via leading-oder $A = 2$ interactions: $A_{4}^{(A = 2)}(1^{+\frac{3}{2}},2^{+\frac{3}{2}},3^{-\frac{3}{2}},4^{-\frac{3}{2}})$. As this is a minimal amplitude, we consider the case where the like-helicity spin-3/2 states are identical: there is only one flavor/species of a spin-3/2 state. How many poles would such an amplitude have? By Eq.~\eqref{Constraint1}, there must be
\eq
2H + 1 - A = N_p \longrightarrow N_p = 2 \, \label{SUGRApoles}
\eqe
poles in any four-point amplitude constructed from $A = 2$ three-point amplitudes which has spin-3/2 states as its highest-spin external state. The key point here is really only that $N_p > 0$: the amplitude must have a factorization channel. Because it has two poles, at least one of them must be mediated by graviton exchange. In this minimal theory, as (a) gravitons can only be produced through particle-antiparticle annihilation channels and (b) the like-helicity spin-3/2 states are identical, both channels occur via graviton exchange. See Fig.~\ref{SUGRAfact}(a) for specifics.

Because this set of external states should always be present in \emph{any} theory with leading-order interactions between any number of spin-3/2 states, S-matrices of these theories must \emph{always} include the graviton. 

This can be made even more explicit. Consider an S-matrix constructed, at least in part, from a three-point amplitude, $A_3(3/2,a,b)$, and its conjugate, $A_3(-3/2,-a,-b)$, where $H = 3/2$ and $A = 2$. These three-point amplitudes tie-together a spin-3/2 state with two other states which, collectively, have helicity-magnitudes $|H| \leq 3/2$. This theory necessarily contains the four-point amplitude,
\eq
A_{4}(1^{+\frac{3}{2}},2^{+a},3^{-\frac{3}{2}},4^{-a}) \, . \label{SUGRA1}
\eqe
As noted in Eq.~\eqref{SUGRApoles}, the denominator within this amplitude has two kinematic poles. Clearly the $s$-channel is has the spin-$b$ state for an intermediary. However, as (a) the three-point amplitudes in the theory all have $A = 2$, and (b) the opposite-helicity spin-3/2 states (equivalently, the spin-$a$ states) are antiparticles, the $u$-channel factorization channel must be mediated by a massless spin-2 state: the graviton. This is depicted in Fig.~\ref{SUGRAfact}. 

Note that the t-channel is also possible, mediated by a helicity $a+1/2$ particle. However, repeating the above reasoning for the new $A_3(3/2,a+1/2,-a)$ amplitude will eventually lead to the necessity of introducing a graviton. This is because in each step the helicity of $a$ is increased by $1/2$, and this process stops once $a$ reaches $3/2$, when both the t and u-channels can only be mediated by a graviton. This pattern of adding particles with incrementally different spin will be investigated further in the following sections. 

Before delving into details of the spectra in theories with multiple species of spin-3/2 states, we note one final feature of these theories. Analysis of their four-particle amplitudes, e.g. the amplitude in Eq.~\eqref{SUGRA1}, via on-shell methods such as the Mandelstam deformation introduced in the previous section, straightforwardly shows that the coupling constants in this theory [the $g^{\pm}_{abc}s$ in the language of Eq.~\eqref{3ptA3}] are \emph{equal} to $\kappa = 1/M_{pl}$, the graviton self-coupling constant. More generally, in any $A = 2$ theory with spin-3/2 states, \emph{each and every} defining three-point amplitude, $A_3(1_a^{h_a},2_b^{h_b},3_c^{h_c}) = \kappa_{abc} M_{abc}(\langle \,,\, \rangle)$ has an identical coupling constant, $\kappa_{abc} = \kappa = 1/M_{pl}$, up to (SUSY preserving Kronecker) delta-functions in flavor-space.

It is important to emphasize here that, as the spin-3/2 gravitinos only interact via $A = 2$ three-point amplitudes, they cannot change the $A < 2$ properties of any state within the same amplitude. Concretely, a bosonic (fermonic) state which transforms under a given specific representation of a compact Lie Algebra, i.e. a particle which interacts with massless vectors (gluons) via leading order ($A = 1$) interactions, can only interact with a fermonic (bosonic) state which transforms under the \emph{same} representation of the Lie Algebra when coupled to spin-3/2 states within $A = 2$ three-point amplitudes. From the point of view of the marginal interactions, only the spin of the states which interact with massless spin-3/2 ``gravitino(s)'' may change. This is the on-shell version of the statement that all states within a given supermultiplet have the same quantum-numbers, but different spins.

We now consider the detailed structure of interactions between states of various different helicities which participate in S-matrices that couple to massless spin-3/2 states. Minimally, such theories include a single graviton and a single spin-3/2 state (and its antiparticle). Equipped with this, we can ask what the next-to-minimal theory might be. There are two ways one may enlarge the theory: (1) introducing a state with a \emph{new} spin into the spectrum of the theory, or (2) introducing another species of massless spin-3/2 state. We pursue each in turn.

\subsection{Minimal extensions of the ${\cal N} = 1$ supergravity theory}\label{SUSYextend}

First, we ask what the minimal enlargement of the ${\cal N} = 1$ SUGRA theory is, if we require inclusion of a single spin-1 vector into the spectrum. In other words, what three-particle amplitudes must be added to,
\eq
{\cal N} = 1 \, {\rm SUGRA} \Longleftrightarrow \big\{ A_3(+2,\pm2,-2)\, , \, A_3(+3/2,\pm2,-3/2) \big\} \, ,
\eqe
in order for all four-particle amplitudes to factorize properly on all possible poles, once vectors are introduced into the spectrum. Clearly, inclusion of a vector requires inclusion of,
\eq
A_3(+1,\pm2,-1) \, ,
\eqe
into the theory. It is useful to consider the four-particle amplitude, $A_4(-3/2,+3/2,+1,-1)$; its external states are only those known from the minimal theory and this extension, i.e. a particle-antiparticle pair of the original spin-3/2 ``gravitino'' and a particle-antiparticle pair of the new massless spin-1 vector. 

By Eq.~\eqref{SUGRApoles}, this amplitude must have two poles. The $s$-channel pole is clearly mediated by graviton-exchange, as the amplitude's external states are composed of two distinct pairs of antiparticles. The \emph{second} pole brings about new states. There are two options for which channel the second pole is associated with: the $u$-channel pole, or the $t$-channel pole. On the $t$-channel, the amplitude must factorize as,
\eq
A_4(-3/2,+3/2,+1,-1)\bigg|_{t \rightarrow 0} \rightarrow A_3(1^{-\frac{3}{2}},4^{-1},P_{14}^{+\frac{1}{2}}) \frac{1}{(p_1 + p_4)^2} A_3(2^{+\frac{3}{2}},3^{+1},-P_{14}^{-\frac{1}{2}}) \, ,
\eqe
and we see that, by virtue of the fact that the three-point amplitudes must have $A = 2$, the new particle introduced into the spectrum is a spin-1/2 fermion. This option corresponds to the CPT-complete spectrum for ${\cal N} = 1$ SUGRA combined with the CPT complete spectrum for ${\cal N} = 1$ SYM. In this case, the full $A = 2$ sector of the theory would be,
\eq
\left\{ A_3(2,2,-2)\, , \, A_3\left(2,\frac{3}{2},-\frac{3}{2}\right)\, , \, A_3\left(2,1,-1\right)\, , \, A_3\left(2,\frac{1}{2},-\frac{1}{2}\right)\, , \, A_3\left(\frac{3}{2},1,-\frac{1}{2}\right)   \right\} \, ,
\eqe
Note that, as the spin-2 and spin-3/2 states interact gravitationally, one can add extra, leading-order $A = 1$ (``gauge'') interactions between the spin-$H \leq 1$---but it is not necessary.

If the second pole is in the $u$-channel, then the amplitude must factorize as,
\eq
A_4(-3/2,+3/2,+1,-1)\bigg|_{u \rightarrow 0} \rightarrow A_3(1^{-\frac{3}{2}},3^{+1},P_{13}^{-\frac{3}{2}}) \frac{1}{(p_1 + p_3)^2} A_3(2^{+\frac{3}{2}},4^{-1},-P_{13}^{+\frac{3}{2}}) \, ,
\eqe
and we see that, by virtue of the fact that the three-point amplitudes must have $A = 2$, the new particle introduced into the spectrum must be \emph{another} spin-3/2 gravitino. This option corresponds to the CPT-complete spectrum for ${\cal N} = 2$ SUGRA.

It is not immediately obvious that this internal spin-3/2 state must be distinguishable from the original spin-3/2 state. Distinguishability comes from the fact that the factorization channel $A_3(1_{a}^{3/2},2_{b}^{3/2},P^{-1}) A_3(3_{\bar{a}}^{-3/2},4_{\bar{b}}^{-3/2},-P^{+1})/K_{12}^2$ must be part of a four-particle amplitude with both the ``new'' and the ``old'' spin-3/2 species as external states. This amplitude also has only two poles. As one of them is mediated by vector exchange, we see that there is only \emph{one} graviton-exchange channel. Therefore the new and old spin-3/2 states cannot be identical.

Constructing theories in this way is instructive. As a consequence of requiring a unitary interpretation of all factorization channels in non-minimal S-matrices involving massless spin-3/2 states, we are forced to introduce a new fermion for every new boson and vice-versa. Further, we see that through allowing minimal extensions to this theory, we can either have extended supergravity theories, i.e. ${\cal N} = 2$ SUGRA theories, truncations of the full ${\cal N} = 8$ SUGRA multiplet, or supergravity theories and supersymmetric Yang-Mills theories in conjunction, i.e. ${\cal N} = 1$ SUGRA $\times \,\, {\cal N} = 1$ SYM theories. The same lessons apply for more extended particle content. However, it is difficult to make such constructions systematic. Below, we discuss the second, more systematic, procedure which hinges upon the existence of ${\cal N}$ distinguishable species of spin-3/2 fermions.

\subsection{Multiple spin-$\frac{3}{2}$ states and (super)multiplets}\label{SUSYconstruct}

Another way to understand these constructions is as follows: specify the number ${\cal N}$ of distinguishable species of spin-3/2 states, and then specify what else (besides the graviton) must be included into the theory. This amounts to specifying the number of supersymmetries and the number and type of representations of the supersymmetry algebra. In the above discussion, the two minimal extensions to the ${\cal N} = 1$ SUGRA theory were: (a) ${\cal N} = 1$ SUGRA $\times \,\, {\cal N} = 1$ SYM, and (b) ${\cal N} = 2$ SUGRA, with \emph{two} gravitinos and one vector. 

To render this construction plan unique, we require that all spins added to the theory besides the graviton and the ${\cal N}$ gravitinos, i.e. all \emph{extra} supermultiplets included in the theory, be the ``top'' helicity component of whatever comes later. So, again, the discussion in subsection~\ref{SUSYextend} would cleanly fall into two pieces: (A) a \emph{single} gravitino (in the graviton supermultiplet) together with a gluon and its descendants, and (B) \emph{two} distinct gravitinos (in the graviton supermultiplet) and their descendants. Clearly this procedure can be easily extended (see subsection~\ref{maxSUSY}).

The general strategy is to look at amplitudes which tie together gravitinos and lower-spin descendants (ascendants) of the ``top'' (``bottom'') helicities in the theory, of the type
\eq
A(1_x^{+3/2},2_y^{-3/2},3_{u}^{-s},4_{v}^{+s})\, . \label{archetype}
\eqe 
Here, the $x$ and $y$ labels describe the species information of the two gravitinos, and $u$ and $v$ describe the species information of the lower-spin particles in the amplitude. Note graviton-exchange can only happen in the $s$-channel. Unitary interpretation of one of the other channels generically forces the existence of a new spin-$s-\frac{1}{2}$ state into the spectrum. For pure SUGRA (i.e. no spin-1, 1/2, or 0 ``matter'' supermultipets), this works as follows,
\begin{enumerate}
\item One gravitino ($\{a\}$). The unique amplitude to consider, after the archetype in Eq.~\eqref{archetype}, is $A_4\big(1_a^{3/2},2_a^{-3/2}, 3_a^{3/2},4_a^{-3/2}\big)$. Both $s$- and $t$-channels occur via graviton-exchange. The theory is self-complete: the other amplitude does not require any new state. 
\item Two gravitinos ($\{a, b\}$). The unique amplitude to consider is $A_4\big(1_a^{3/2},2_a^{-3/2}, 3_b^{3/2},4_b^{-3/2}\big)$. Here, the $t$-channel is disallowed; the $u$-channel needs a vector with gravitino-label $\{ab\}$. Inclusion of this state completes the theory.
\item Three gravitinos ($\{a, b, c\}$). Two classes amplitudes of the type in Eq.~\eqref{archetype} to consider. First, $A_4\big(1_a^{3/2},2_a^{-3/2}, 3_c^{3/2},4_c^{-3/2}\big)$ requires a vector with gravitino-label $ac$ in its $u$-channel; as there are three amplitudes of this type, there are three distinguishable vectors: $\{ab, ac, bc\}$. Second, $A_4\big(1_a^{3/2},2_a^{-3/2}, 3_{bc}^{+1},4_{bc}^{-1}\big)$ needs a fermion with gravitino-label $\{abc\}$ in the $u$-channel. No other amplitudes require any new states.
\item Four gravitinos ($\{a, b, c, d\}$). Here, the structure is slightly more complicated, but similarly hierarchical. Three classes of amplitudes, each following from its predecessor. First, there are $4\choose2$ distinct $A_4\big(1_a^{3/2},2_a^{-3/2}, 3_c^{3/2},4_c^{-3/2}\big)$s. They require vectors with gravitino labels $\{ab, ac, ad, bc, bd, cd\}$. Second, there are $4\choose3$ distinct $A_4\big(1_a^{3/2},2_a^{-3/2}, 3_{bc}^{+1},4_{bc}^{-1}\big)$s, which require spin-1/2 fermions with labels $\{ abc, abd, acd, bcd\}$. Third and finally, we consider $A_4\big(1_a^{3/2},2_a^{-3/2}, 3_{bcd}^{1/2},4_{bcd}^{-1/2}\big)$. On its $u$-channel, it requires a spin-0 state with gravitino-label $\{abcd\}$.
\end{enumerate}
Crucially, we observe that all spins present in the graviton supermultiplet (the graviton and all of its descendants) with ${\cal N}$ gravitinos are still present in the graviton supermultiplet with ${\cal N} + 1$ gravitinos---but with higher multiplicities. These descendant states are explicitly labeled by the gravitino species from whence they came. Spin-$s$ states in the graviton multiplet have ${\cal N}\choose s$ distinct gravitino/SUSY labels. 

Importantly, if $h$ is the unique lowest helicity descendant of the graviton with ${\cal N}$ gravitinos, then inclusion of an extra gravitino allows for ${\cal N}$ \emph{new} helicity-$h$ descendants of the graviton. Now, studying $A_4\big(1_{{\cal N}+1}^{3/2}, 2_{{\cal N}+1}^{-3/2},3_{ab...{\cal N}}^{+h},4_{ab...{\cal N}}^{-h}\big)$, we see that again the $u$-channel requires a single new descendant with helicity $h-1/2$ and SUSY-label $\{ab...{\cal N},{\cal N}+1\}$. 

This logic holds for the descendants of all ``top'' helicity states: isomorphic tests and constructions, for example, allow one to construct and count the descendants from the gluons of SYM theories. We see below that, by obeying the consistency conditions derived from pole-counting and summarized in Fig.~\ref{HA1}, this places strong constraints on the number of distinct gravitinos in gravitational and mixed gravitational and $A<2$-theories.

\subsection{Supersymmetry, locality, and unitarity: tension and constraints}\label{maxSUSY}

As we have seen, inclusion of ${\cal N}$ distinguishable species of massless spin-$3/2$ states into the spectrum of constructible theories forces particle helicities $\{H, H-1/2, ... , H - {\cal N}/2\}$ into the spectrum. But, as we have seen in sections~\ref{PoleCount} and~\ref{UniqueLaws}, the $A =2$ gravitational interactions cannot consistently couple to helicities $|h| > 2$. And so, within the supersymmetric gravitational sector, we must have (a) $H = 2$, and (b) $H - {\cal N}/2 \geq -2$. Otherwise, we must couple a spin-$5/2 > 2$ to gravity--which is impossible. Locality and unitarity constrains ${\cal N} \leq 8$. 

So there is tension between locality, unitarity, and supersymmetry. We now ask about the spectrum of next-to-minimal theories coupled to spin-3/2 states. There are two options for such next-to-minimal theories: either (relevant) self-interacting scalars or (marginal) self-interacting vectors coupled to ${\cal N}$ flavors of spin-3/2 particles. Immediately, we see that $\phi^3$ cannot be consistently coupled to spin-3/2 states. Coupling the spin-0 lines in $\phi^3$ to even one spin-3/2 state would force the existence of non-zero $A_{3}(1/2,-1/2,0)$-type interactions. But these interactions, as discussed in section~\ref{weirdtheory}, are not consistent with unitarity and locality. So \emph{relevant} interactions cannot be supersymmetrized in flat, four-dimensional, Minkowski space.

However, for ($A = 1$) self-interacting gluons, the story is different. By the arguments above, unitarity and locality dictate that if ${\cal N} $ spin-3/2 particles are coupled to gluons, then gluons must couple via marginal interactions, to spin-$\pm|1-{\cal N}/2|$ states. Again basic pole-counting in section~\ref{Big3}, $A = 1$ interactions are only valid for $|h| \leq 1$. And so, we must have ${\cal N} \leq 4$, if we would like to couple interacting vectors to multiple distinct spin-3/2 particles while also respecting locality and unitarity of the S-matrix.

\section{Future directions and concluding remarks}\label{Conclusion}

Our results can be roughly separated into two categories. First, we classify and systematically analyze all possible three-point massless S-matrix elements in four-dimensions, via basic pole-counting. The results of this analysis are succinctly presented in Fig.~\ref{HA1}. Second, we study the couplings and spectra of the few, special, self-interactions allowed by this first, broader-brush, analysis. In this portion of the paper, we reproduce standard results on the structures of massless S-matrices involving higher-spin particles, ranging from the classic Weinberg-Witten theorem and the Equivalence Principle to the existence of supersymmetry, as consequences of consistency conditions on various S-matrix elements. We recap the main results below.

Locality and constructibility fix the generic pole-structure of four-point tree-amplitudes constructed from fundamental higher-spin three-point massless amplitudes. Tension between the number of poles mandated by these two principles, and unitarity, which bounds the number of poles in an amplitude from above ($N_p \leq 3$), eliminates all but a small (yet infinite) sub-class of three-point amplitudes as leading to four-particle tree-level S-matrices that are inconsistent with locality and unitarity.

Already from this point of view we see that, for low $A = |\sum_{i = 1}^3 h_i|$, (Super-)Gravity, (Super-)Yang-Mills, and $\phi^3$-theory are the unique, leading, interactions between particles of spin-$|h| \leq 2$. Further, we see that gravitational interactions cannot \emph{directly} couple to particles of spin-$|H| > 2$. Similarly, massless vectors interacting at leading-order ($A = 1$) cannot consistently couple to massless states with helicity-$|H| > 1$. 

In light of these constraints, we study higher-$A$ theories. The upper-bound on the number of poles in four-particle amplitudes, imposed by unitarity and locality, is even stronger for higher-spin \emph{self-interactions} ($N_p \leq 1$ for $A > 2$). The set of consistent three-point amplitudes with $A > 2$ is further reduced to lie between the lines $H = A/2$ and $H = A/3$. 

Exploiting this, we re-examine whether-or-not the primitive amplitudes which define the S-matrices of General Relativity and Yang-Mills can \emph{indirectly} couple to higher-spin states in a consistent manner. As they cannot directly couple to higher-spin states, this coupling can only happen within non-primitive four-point (and higher) amplitudes, which factorize into GR/YM self-interaction amplitudes (with $H = A$), multiplied by an $A > 2$ three-point amplitude with an external tensor or vector. Again, simple pole-counting shows that amplitudes which couple the $A \le 2$ to the $A > 2$ theories generically have poles whose unitary interpretation mandates existence of a particle with spin $\tilde{H} > A/2$. 

Having such a high-spin particle contradicts the most basic constraint~\eqref{Constraint1}, and thus invalidates the interactions---save for two special examples. These examples are simply the higher-derivative amplitudes which also couple three like-helicity gravitons, $A_3(2,2,2)$, and/or like-helicity gluons, $A_3(1,1,1)$. There is a qualitative difference between massless spin-2(1) particles participating in lower-spin ($A \leq 2$) amplitudes, and massless spin-2(1) particles participating in higher-spin ($A>3$) amplitudes. The graviton is unique. Gluons are also unique. They cannot be coupled to particles of spin-$|h| > 2$!

Equipped with the (now) \emph{finite} list of leading interactions between spin-1, spin-2, and lower-spin states, we then analyze the structure of their interactions---i.e. their coupling constants. To do this, we set up, and show the validity of, the Mandelstam shift~\eqref{defMand1}. Assuming parity-invariance, and thus $g^{+}_{ijk} = + g^{-}_{ijk}$ [in the notation of Eq.~\eqref{3ptA3}], we perform the Mandelstam shift on four-point amplitudes in these theories. Invariance with respect to redefinitions of the unphysical shift-parameter directly implies the Lie Algebraic structure of the marginal ($A = 1$) coupling to massless vectors; similarly massless tensors must couple (a) diagonally (in flavor space), and (b) with equal strength to all states.

Finally, we analyze consistency conditions on four-point amplitudes which couple to massless spin-3/2 states. The minimal theory/set of interacting states, at leading order, which include a single spin-3/2 state is the theory with a single graviton and a single spin-3/2 state, at $A = 2$. From this observation, we identify the spin-3/2 state with the gravitino. The gravitino also couples to matter with strength $\kappa = 1/M_{pl}$, but as it is not a boson, it does not couple ``diagonally'': coupling to non-graviton states within the leading-order $A = 2$ interactions automatically necessitates introduction of a fermion for every boson already present in the theory, and vice-versa. We recover the usual supersymmetry constraints, such as fermion-boson level matching, and the maximal amount of distinguishable gravitinos which may couple to gluons and/or gravitons; above these bounds, the theories becomes inconsistent with locality and unitarity.

We close with future directions. Clearly, it would be interesting to discuss on-shell consistency conditions for theories which have primitive amplitudes which begin at \emph{four} points, rather than at three-points. Certain higher-derivative theories, such as the nonlinear sigma model\cite{Sigma1}, are examples of this type of theory: in the on-shell language, derivative interactions between scalars can only act to give factors of non-trivial kinematical invariants within the numerator of a given amplitude. All kinematical invariants are identically zero at three points. So the first non-zero S-matrix elements in derivatively-coupled scalar theories must be at four-points. Supported by the existence of semi- on-shell recursions in these theories\cite{Sigma2}, it is conceivable that these theories are themselves constructible. Straightforwardly, this leads to the on-shell conclusion that all S-matrix elements in these theories have an even number of external legs. Much more could be said, and is left to future work.

Besides theories with derivative interactions, there is also a large of class of higher-spin theories {\it not} constrained by any of the arguments presented in this paper. These are $A\ge 3$, $N_p\le 1$ theories which do not contain any spin-1 or spin-2 states, for example $A_3(3/2,3/2,0)$. It is not clear from this on-shell perspective whether such theories are completely compatible with locality and unitarity, or more sophisticated tests can still rule them out.

Indeed, an exhaustive proof of the spin-statistics theorem has yet to be produced through exclusively on-shell methods. Proof of this theorem usually occurs, within local formulations of field theory, through requiring no information propagation outside of the light-cone. In the manifestly on-shell formalism, all lines are on their respective light-cones; superluminal propagation, and (micro-)causality violations are naively inaccessible. Ideally some clever residue theorem, such as that in~\cite{Superluminal}, should prove the spin-statistics theorem in one fell swoop. Further, it would be interesting to prove that parity-violation, with $g^{+}_{abc} = - g^{-}_{abc}$, within three-point amplitudes only leads to consistent four-particle amplitudes for parity-violating gluon/photon-fermion amplitudes ($A = 1$). 

Further, one may reasonably ask what the corresponding analysis would yield for \emph{massive} states in constructible theories. As is well known, massive vectors must be coupled to spinless bosons (such as the Higgs), to retain unitarity at $E_{\rm CM} \sim s \gtrsim m_{V}$~\cite{PITP}. It would be extremely interesting to see this consistency condition, and analogous consistency conditions for higher-spin massive particles, naturally fall out from manifestly on-shell analyses.

Finally, one may wonder whether-or-not similar analysis to that presented in this paper could apply to loop-level amplitudes in massless theories. In the on-shell language, loops and trees have varying degrees of transcendental dependence on kinematical invariants. Concretely, tree amplitudes have at most simple poles. However, loop amplitudes have both polylogarithms which have branch-cuts and are functions of ratios of kinematic invariants, and rational terms with higher-order poles in the kinematical invariants, the existence of which is (almost) solely to cancel the higher-order poles in these same kinematic invariants arising from these polylogarithms. See, for example, the discussion in~\cite{Anomalies}~\cite{AnomaliesII}. These branch-cuts and higher-order poles would dramatically complicate any attempt to use the reasoning championed in this paper, at loop-level. Nonetheless, there is a very well-known consistency condition which arises between the interference of tree- and loop-amplitudes: the classic Green-Schwarz anomaly cancellation mechanism! It would be very interesting to pursue, exhaustively, the extent to which similarly inconsistent-seeming tree-amplitudes could be made consistent upon including loops. But this is left for future work.

In conclusion, these results confirm the Coleman-Mandula and the Haag-Lopuszanski-Sohnius theorems for exclusively massless states in four-dimensions~\cite{CM}\cite{HLS}. Through assuming a constructible, non-trivial, S-matrix that is compatible with locality and unitarity, we see that the maximal structure of non-gravitational interactions between low-spin particles is that of compact Lie groups. Only through coupling to gravitons and gravitinos can additional structure be given to the massless tree-level S-matrix (at four-points). This additional structure is simply supersymmetry; it relates scattering amplitudes with asymptotic states of different spin, within the same theory. Further, no gravitational, marginal, or relevant interaction may consistently couple to massless asymptotic states with spin greater than two.

\section*{Acknowledgements}

The authors thank Nima Arkani-Hamed for suggesting this topic, and for his continual encouragement and guidance throughout the process. We thank Clifford Cheung, Yu-tin Huang, David Skinner, and Jaroslav Trnka, and Alexander Zhiboedov for helpful and encouraging discussions on various topics. Finally, we thank Stanley Deser, Sylvester James Gates, Massimo Porrati, Augusto Sagnotti, Mirian Tsulaia and collaborators, for helpful discussions of previous, related, work.

\appendix

\section{Constructing minimal numerators}\label{n4construct}

Here, we prove that the minimal numerator for a four-point amplitude of massless particles satisfies Eq.~\eqref{n4anatomy1} for the special case where the sum of all four helicities vanishes.\footnote{All four-point tree-amplitudes constructed from a set of three-point amplitudes and their conjugate amplitudes, $A_3 \ \& \ \bar{A}_3$, have this property.}
Given $A_4(1^{h_1},2^{h_2},3^{h_3},4^{h_4})$, we re-label the external states by increasing helicity: 
\begin{align}
H_1 \geq H_2 \geq H_3 \geq H_4
\label{ordering}
\end{align}  
The total helicity vanishes, and thus $H_1 \geq 0$ and $H_4 \leq 0$.  Now, define $H_1^{+} = |H_1|$ and $H_4^{-} = |H_4|$: the numerator has at least $2H_1^{+} \ \tilde{\lambda}$s, and at least $2H_4^{-} \ \lambda$s.

Now there must exist 
\begin{align}
N_{\tilde{\lambda}}^{\rm ext.} = 2H_1^{+} + N_{\tilde{\lambda}}^{\rm rest} \geq 2 H_1^+
\label{Hplus}
\end{align}
total external $\tilde{\lambda}$s in the numerator.  Similarly, the numerator must contain a total of
\begin{align}
N_{\lambda}^{\rm ext.} = 2H_4^{-} + N_{\lambda}^{\rm rest} \geq 2 H_4^-
\label{Hminus}
\end{align}
external $\lambda$s.

By definition, the numerator is both (a) Lorentz-invariant, and (b) little-group covariant, and (c) encodes all of the helicity information of the asymptotic scattering states.  Therefore, it must be of the form
\begin{align}
{\rm Numerator} \sim \langle \, \rangle_{(1)} ... \langle \, \rangle_{(n)} \,\,\, [ \, ]_{(1)} ... [ \, ]_{(m)} \ .
\end{align}  
Notably, requiring $\sum^4_{a = 1} H_a = 0$ directly implies that the numerator contains equal number of holomorphic and anti-holomorphic spinor-helicity variables, 
\begin{align}
\sum^4_{a = 1} H_a = 0 \Leftrightarrow \big\{ N^{\rm ext.}_{\lambda} = N^{\rm ext.}_{\tilde{\lambda}} \ , \ {\rm and} \ N_{\lambda}^{\rm total} = N_{\tilde{\lambda}}^{\rm total} \big\}
\end{align}

We note that because the numerator contains the same number of $\lambda$s as $\tilde{\lambda}$s, and must be a product of spinor-brackets, then it must have the same number of each type of spinor-product: 
\begin{align}
{\cal N}_b = N_{\langle \ \rangle} = N_{[ \ ]} \ . 
\label{NB}
\end{align}
The reason is as follows. First, note that only inner-products of spinor-helicity variables are both (a) Lorentz-invariant and (b) little-group covariant.  Because the numerator has both of these properties, all of the $\lambda$s and $\tilde{\lambda}$s which encapsulate the helicity information of the asymptotic scattering states must be placed within spinor-brackets.  If we take $N_{\langle \ \rangle} \neq N_{[ \ ]}$, then there would be a mis-match between the number of $\lambda$s and $\tilde{\lambda}$s in the numerators. This contradicts the statement that the numerator must contain the same number of positive- and negative- chirality spinor-helicity variables. This proves Eq.~\eqref{NB}.

Now, the \emph{minimal} number of spinor-brackets ${\cal N}$ is simply given by 
\begin{align}
{\cal N}_b = 2 \times {\rm max} \{ H_1^+, H_4^-\}
\label{NB2}
\end{align}
This can be seen in the following way. At the minimum, there must be $2H_1^+ \ [ \ ]$s and $2 H_4^- \ \langle \ \rangle$s within the numerator. Otherwise, at least two of the $2H_1^+$ copies of $\tilde{\lambda}_1$ within the numerator would have to within the same spinor bracket, $[\tilde{\lambda}_1,\tilde{\lambda}_1]$. But this would force the numerator to vanish. As we are only concerned with non-trivial amplitudes, we thus require $N_{[ \ ]} \geq 2 H_1^+$. The same logic requires $N_{\langle \ \rangle} \geq 2 H_4^-$.

But, by Eq.~\eqref{NB}, we must have $N_{[ \ ]} = N_{\langle \ \rangle}$.  So we must have ${\cal N}_b \geq 2 \times {\rm max} \{H_1^+, H_4^-\}$. The minimal numerator saturates this inequality. This proves Eq.~\eqref{NB2}.

It is important to note that this minimal number of spinor-brackets of each type, ${\cal N}_b = 2 {\rm max} \{H_1^+, H_4^-\}$, mandated by Eq.~\eqref{NB2} to be present within the numerator is ``large'' enough to encode the helicity information of all of the external scattering states---not just the helicity information of $1^{+H_1^+}$ and $4^{-H_4^-}$. 

In other words, there are ``enough'' $\langle \ \rangle$s and [ \ ]s already present in the numerator to fit in the remaining $\lambda$s and $\tilde{\lambda}$s required to encode the helicity information of the other two particles. I.e.,
\begin{align}
N_{[ \ ]} \geq N_{\tilde{\lambda}}^{\rm rest} \ , \ {\rm and} \ N_{\langle \ \rangle} \geq N_{\lambda}^{\rm rest} 
\label{sufficient}
\end{align}
Before proving this, first recall eqs.~\eqref{Hplus},~\eqref{Hminus}, and~\eqref{NB2}. By~\eqref{NB2}, $N_{[ \ ] } = N_{\langle \ \rangle} = 2 {\rm max} \{H_1^+,H_4^-\}$. Now, how many $\lambda$s and $\tilde{\lambda}$s must be present in the numerator to ensure all external helicity data is properly entered into the numerator? There are only three cases to consider. In all cases, Eq.~\eqref{sufficient} holds:
\begin{enumerate}
\item Particles 1 and 2 have positive helicity, while particles 3 and 4 have negative helicity. Now, by definition, we would like to show that of the $2 {\rm max} \{H_1^+,H_4^-\} \ [ \ ]$s required by Eq.~\eqref{NB2} are sufficiently numerous to allow inclusion of $2|H_2|$ more $\tilde{\lambda}_2$s. This is guaranteed by the orderings: $H_1 \geq H_2$. So there are enough empty slots in the anti-holomorphic spinor-brackets to encode the helicity of all positive-helicity particles. The same holds for $H_3$.   For this case, Eq.~\eqref{sufficient} holds.
\item Only particle 4 has negative helicity. All others have positive helicity. Because the amplitudes under consideration have total helicity zero, we know that the sum of helicities of the particles with positive helicity must equal $H_4^-$. Hence there must be $2H_4^- \lambda_4$s and $2 H_4^-$ physical $\tilde{\lambda}$s in the numerator. Further, as only particle 4 has negative helicity, it follows that ${\cal N}_b = 2{\rm max}\{H_1^+, H_4^- \} = 2 H_4^{-}$. And so there are $2 H_4^{-}$ spinor-brackets of each kind. For this case Eq.~\eqref{sufficient} holds.
\item Only particle 1 has positive helicity. This case is logically equivalent to the above. 
\end{enumerate}
This proves Eq.~\eqref{sufficient}, and therefore proves Eq.~\eqref{n4anatomy1}:
\begin{align}
N \sim \langle \, \rangle_{(1)} ... \langle \, \rangle_{(2H)} \,\,\, [ \, ]_{(1)} ... [ \, ]_{(2H)}  \Rightarrow [N] = (K^2)^{2H} \ , \end{align}
where $H = {\rm max} \{|h_1|,...,|h_4|\}$.  Establishing this result concludes the proof.

\section{Ruling out theories with $N_p = 2$, for $A \geq 3$}\label{Np2explicit}

In this section, we rule out self-interacting theories constructed from three-point amplitudes which necessitate two poles in the four-point amplitudes, for $A > 2$. This is simple pole-counting, augmented by constraint~\eqref{Constraint1} and the results of subsection~\ref{Np32} for $N_p = 3$. Recall, for amplitudes within a self-interacting sector of a theory, we have $\sum_{i = 1}^4 h_{i}^{\rm ext} = 0$, and $N_p = 2H + 1 - A$. For $N_p = 2$, we must have $H = {\rm max}\{ |f|, |H|, |g| \} = (A+1)/2$. 

Within this sector we may construct $A_4(1^{+H},2^{-H},3^{+f},4^{-f})$ from $A_3(H, g, f)$. By assumption, it has two factorization channels, specifically the $t$- and $u$-channels.\footnote{The $s$-channel in this amplitude is disallowed, as it would require a new particle with helicity $\tilde{H} = \pm A$, which would lead to amplitudes with $N_p  = A+1$.} Without loss of generality, we take $f > 0$. The intermediary in the $u$-channel pole has spin $g = A - (H+f) = (A-1)/2 - f < H = (A+1)/2$, and poses no barrier to a unitary \& local interpretation of the amplitude/theory.

However, for the $t$-channel's intermediary must have helicity $\tilde{H} = A + f - H = (A-1)/2 + f$. And so, for $f > 1$, we must include a new state with larger helicity $\tilde{H} = H + (f-1) > H$. As discussed in subsection~\ref{Np32}, this does not a priori spell doom for the theory. However, in this case it does: inclusion of this new, larger helicity, state within the theory forces inclusion of four-point amplitudes with these states on external lines.  These \emph{new} amplitudes have a larger number of poles: $2\tilde{h} + 1 - A = (2H+1-A) + 2f = 2 + 2f \geq 4 > 3$, for $f > 1$. 

Theories with $H = (A+1)/2$, and $f > 1$ (or $g > 1$) cannot be consistent with unitarity and locality. Inspection reveals that all theories with $N_p = 2$ and $A > 2$ are of this type, save for three special examples. Explicitly, for $A = 3, \,4, \,5$, and $A = 6$, the $N_p=2$ theories are defined by $(H,g,f)$s of the following types,
\begin{align}
A = 3&:\quad  \left(2,2,-1 \right) \, , \, \left(2,\frac{3}{2},-\frac{1}{2} \right) \, , \, \left(2,1,0 \right) \, , \, \left(2,\frac{1}{2},\frac{1}{2} \right) \, , \label{B1}\\
A = 4&:\quad  \left(\frac{5}{2},\frac{5}{2},-1 \right) \, , \, \left(\frac{5}{2},2,-\frac{1}{2} \right) \, , \, \left(\frac{5}{2},\frac{3}{2},0 \right) \, , \, \left(\frac{5}{2},1,\frac{1}{2} \right) \, , \label{B2}\\
A = 5&:\quad  \left(3,3,-1 \right) \, , \, \left(3,\frac{5}{2},-\frac{1}{2} \right) \, , \, \left(3,2,0 \right) \, , \, \left(3,\frac{3}{2},\frac{1}{2} \right) \, , \, \left(3, 1, 1 \right) \, , \\
A = 6&:\quad  \left(\frac{7}{2},\frac{7}{2},-1 \right) \, , \, \left(\frac{7}{2},3,-\frac{1}{2} \right) \, , \, \left(\frac{7}{2},\frac{5}{2},0 \right) \, , \, \left(\frac{7}{2},2,\frac{1}{2} \right) \, , \, \left(\frac{7}{2}, \frac{3}{2}, 1 \right) \, ,
\end{align}
and their conjugate amplitudes. Clearly, all but the last two entries on line~\eqref{B1} and the last entry on line~\eqref{B2} have $f > 1$ (thus $\tilde{H} > (A+1)/2$), and are inconsistent. Further, it is clear that all higher-$A$ $N_p = 2$ theories may only have pathological three-point amplitudes, which indirectly lead to this same tension with locality and unitarity: except for those three special cases, all $N_p = 2$ theories must have $f$s that are larger than unity.

It is a simple exercise to show that these three pathological examples are inconsistent: playing around with the factorization channels of $A_4(H,-H,f,-f)$s reveals that again the $t$-channel is the problem. The $t$-channel requires the three-point amplitudes on lines~\eqref{B1} or~\eqref{B2} which are \emph{directly} ruled out, as they have $H = (A+1)/2$ and $f > 1$. 

No three-point amplitude with $N_p = 2H + 1 - A = 2$ can lead to a constructible S-matrix consistent with locality and unitarity, for any $A$ larger than two.

\section{Uniqueness of spin-3/2 states}\label{UniqueSUGRA}

In this appendix, we will use similar arguments to those in section~\ref{UniqueLaws} to show that massless spin-3/2 states can only couple consistently to massless particles with helicities $|H| \leq 2$. Recall that for $A > 2$, no constructible theory can be consistent with unitarity and locality unless $A/3 \leq H \leq A/2$. To see whether-or-not the gravitino discussed in section~\ref{SUSY} can couple to any higher-$A$ amplitude, we simply study four-point amplitudes with factorization channels of the type,
\eq
A_4(1^{+2},2^{-\frac{3}{2}},3^{-c},4^{-d}) \rightarrow A_3^{(GR)}(1^{+2},2^{-\frac{3}{2}},P^{+\frac{3}{2}}) \frac{1}{s} A_3^{(A)}(P^{-\frac{3}{2}},3^{-c},4^{-d}) \, .
\eqe
Note that only two three-point amplitudes are consistent with the dual requirements (a) $A/3 \leq H \leq A/2$ and (b) $3/2 \in \{ h_1, h_2, h_3\}$ for $A > 2$. These theories, and their corresponding four-particle amplitudes, are:
\begin{align}
A_3(A/2-1/2, 3/2, A/2-1) &\Rightarrow A_4(1^{+2},2^{-\frac{3}{2}},3^{-(A/2-1/2)},4^{-(A/2 -1)}) \, , \,\, {\rm and} \nonumber\\
A_3(A/2, 3/2, A/2-3/2) &\Rightarrow A_4(1^{+2},2^{-\frac{3}{2}},3^{-A/2},4^{-(A/2 -3/2)}) \, .
\end{align}
Now, the minimal numerator which encodes the helicity information of, for instance, the first amplitude is,
\eq
N \sim [1| P | 2 \rangle^3 [1| Q | 3 \rangle \big(\langle 3, 4 \rangle^2\big)^{A/2-1} \Rightarrow [N] = \big(K^2\big)^{(A/2)+3} \, .
\eqe
However, by power-counting, the kinematic part of the amplitude must have mass-dimension,
\eq
\bigg[ \frac{N}{f(s,t,u)} \bigg] = \bigg[ \frac{1}{K^2} A_{\rm Left}^{(GR)} A_{\rm Right}^{(A)} \bigg] = \frac{(K^2)^{1/2} (K^2)^{A/2}}{(K^2)} = (K^2)^{A/2} \, ,
\eqe
and thus the denominator, $f(s,t,u)$ must have mass-dimension three:
\eq
[f(s,t,u)] = (K^2)^3 \Rightarrow f(s,t,u) = s \, t \, u \, !
\eqe
However, as is obvious from inspection of any amplitude for $A \geq 4$, two of these factorization channels require inclusion of states with helicities which violate the most basic constraint~\eqref{Constraint1}. Thus, no spin-3/2 state in any three-point amplitude with $A \geq 4$ can be identified with the gravitino of section~\ref{SUSY}. The sole exception to this is the three-point amplitude $(H,A) = (3/2, 3)$:
\eq
A_3(1^0,2^{+\frac{3}{2}},3^{+\frac{3}{2}}) \Rightarrow A_4(1^{+2},2^{-\frac{3}{2}},3^{-\frac{3}{2}},4^{0}) \, .
\eqe
Factorization channels in this putative amplitude necessitate only either scalar or gravitino exchange, and are thus not in obvious violation of the consistency condition~\eqref{Constraint1}.

\section{$F^3$- and $R^3$-theories and SUSY}\label{SUGRAexclude}

Basic counting arguments show us that the $F^3$- and $R^3$- theories, i.e. the S-matrices constructed from $A_3(1,1,1)$ and $A_3(2,2,2)$ and their conjugates, are not compatible with leading-order (SUGRA) interactions with spin-3/2 states. The argument is simple.

First, we show that $F^3$-theories are not supersymmetrizable. Begin by including the minimal ${\cal N} = 1$ SUGRA states, together with the three-particle amplitudes which couple gluons to the single species of spin-3/2 (gravitino) state that construct the ${\cal N} = 1$ SYM multiplet. Additionally, allow the $F^3$-three-point amplitude as a building-block of the S-matrix. In other words, begin consider the four-particle S-matrix constructed from,
\eq
A_3\left(2,\frac{3}{2},-\frac{3}{2} \right) \, , \,\, A_3\left(\frac{3}{2},1,-\frac{1}{2} \right) \, ,\,\, A_3\left(1 ,\frac{1}{2},-\frac{3}{2} \right) \, , \,\, {\rm and} \,\, A_3\left(1,1,1 \right) \, , \label{F3susy}
\eqe
where all spin-1 states are gluons, and all spin-1/2 states are gluinos. Now, consider the four-particle amplitude, $A_4(+\frac{3}{2},-\frac{1}{2},-1,-1)$. On the $s$-channel, it factorizes nicely:
\eq
A_4\left(1^{+\frac{3}{2}},2^{-\frac{1}{2}},3^{-1},4^{-1}\right)\bigg|_{s \rightarrow 0} 
\rightarrow \frac{1}{s} 
A_3\left(1^{\frac{3}{2}},2^{-\frac{1}{2}},P^{+1}\right)
A_3\left(P^{-1},3^{-1},4^{-1}\right) \, . \label{F3amp}
\eqe
Clearly, it fits into the theory defined in Eq.~\eqref{F3susy}. To proceed further, we note that its minimal numerator must have the form,
\eq
N \sim [1| P | 2 \rangle [1| Q | 3 \rangle [1| K | 4 \rangle \langle 3 4 \rangle , \label{F3num} \, .
\eqe
Now, this amplitude must have kinematic mass-dimension,
\eq
\left[ \frac{A_3^{\rm SUGRA} A_{3}^{F^3}}{K^2}\right] = \left[ \frac{[ \,\, ]^2 \langle \,\, \rangle^3}{\langle \rangle [ \,\,] } \right] = \left( K^2 \right)^{1+\frac{1}{2}} \label{F3mass} \, .
\eqe
Combining Eq.~\eqref{F3num} and Eq.~\eqref{F3mass}, we see that $1/f(s,t,u)$ must have two poles. On the \emph{other} pole, say on the $t \rightarrow 0$ pole, it takes the form
\eq
A_4\left(1^{+\frac{3}{2}},2^{-\frac{1}{2}},3^{-1},4^{-1}\right)\bigg|_{t \rightarrow 0} 
\rightarrow \frac{1}{t} 
A_3\left(1^{\frac{3}{2}},4^{-1},P^{+\Delta}\right)
A_3\left(P^{-\Delta},2^{-\frac{1}{2}},3^{-1}\right) \, .
\label{F3wrong}
\eqe
Now, one of these two sub-amplitudes must have $A = 3$. However, recall that in section~\ref{UniqueLaws} we showed that the only three-point amplitude which may have spin-1 states identified with the gluons is the $A_3(1,1,1)$ amplitude. Observe that, regardless of which amplitude has $A = 3$, \emph{both} amplitudes contain one spin-1 state and another state with spin-$s \neq 1$. Therefore \emph{neither} amplitude can consistently couple to the $A = 1$ gluons. Therefore we conclude that ${\cal N} = 1$ supersymmetry is incompatible with $F^3$-type interactions amongst gluons.

Similar arguments show that the three-point amplitudes arising from $R^3$-type interactions cannot lead to consistent S-matrices, once spin-3/2 gravitinos are included in the spectrum. Again, we first specify the four-particle S-matrix as constructed from the following primitive three-particle amplitudes:
\eq
A_3\left(2,\frac{3}{2},-\frac{3}{2} \right) \, , \,\, {\rm and} \,\, A_3\left(2,2,2 \right) \, , \label{R3susy}
\eqe
Now, consider the four-particle amplitude, $A_4(+\frac{3}{2},-\frac{3}{2},-2,-2)$, an analog to that considered in the $F^3$-discussion. On the $s$-channel, it factorizes nicely:
\eq
A_4\left(1^{+\frac{3}{2}},2^{-\frac{3}{2}},3^{-2},4^{-2}\right)\bigg|_{s \rightarrow 0} 
\rightarrow \frac{1}{s} 
A_3\left(1^{\frac{3}{2}},2^{-\frac{3}{2}},P^{+2}\right)
A_3\left(P^{-2},3^{-2},4^{-2}\right) \, . \label{R3amp}
\eqe
Clearly, it fits into the theory defined in Eq.~\eqref{R3susy}. To proceed further, we note that its minimal numerator must have the form,
\eq
N \sim [1| P | 2 \rangle^3 \left(\langle 3 4 \rangle^2\right)^2 \label{R3num} \, .
\eqe
Now, this new amplitude must have kinematic mass-dimension,
\eq
\left[ \frac{A_3^{\rm SUGRA} A_{3}^{R^3}}{K^2}\right] = \left[ \frac{[ \,\, ]^2 \langle \,\, \rangle^6}{\langle \rangle [ \,\,] } \right] = \left( K^2 \right)^{3} \label{R3mass} \, .
\eqe
Combining Eq.~\eqref{R3num} and Eq.~\eqref{R3mass}, we see that $1/f(s,t,u)$ must have two poles, again, as in the $F^3$-discussion above. On the \emph{other} pole, say on the $t \rightarrow 0$ pole, it takes the form
\eq
A_4\left(1^{+\frac{3}{2}},2^{-\frac{3}{2}},3^{-2},4^{-2}\right)\bigg|_{t \rightarrow 0} 
\rightarrow \frac{1}{t} 
A_3\left(1^{\frac{3}{2}},4^{-2},P^{+\Delta}\right)
A_3\left(P^{-\Delta},2^{-\frac{3}{2}},3^{-2}\right) \, .
\label{R3wrong}
\eqe
Reasoning isomorphic to that which disallowed ${\cal N} = 1$ SUSY and $F^3$-gluonic interactions rules out the compatibility of this given factorization channel with ${\cal N} = 1$ SUSY and $R^3$-effective gravitational interactions. Namely, it must be that one of these two sub-amplitudes must have $A = 6$. However, recall that in section~\ref{UniqueLaws} we showed that the only three-point amplitude which may have spin-1 states identified with the gluons is the $A_3(1,1,1)$ amplitude. Observe that, regardless of which amplitude has $A = 6$, \emph{both} amplitudes contain one spin-1 state and another state with spin-$s \neq 2$. Therefore \emph{neither} amplitude can consistently couple to the $A = 2$ gravitons. Therefore we conclude that ${\cal N} = 1$ supersymmetry is \emph{also} incompatible with $R^3$-type interactions amongst gravitons.

\section{Justifying the complex deformation in section~\ref{niceshift}}\label{Illustration}

One might worry about the validity of such a shift, and how it could be realized in practice. In other words, one could wonder whether-or-not shifting the Mandlsetam invariants, $(s,t,u) \rightarrow (s+z \tilde{s},t+z \tilde{t},u+z \tilde{u})$ would not also shift the numerator of the amplitude. Here, we prove that such a shift must always exist.

First, a concrete example. Suppose one desired to study the constraints on the $f_{abc}$s characterizing $A_3(1_a^{+1},2_b^{-1},3_c^{-1}) = f_{abc} \langle 2 3 \rangle^3/\{\langle 31 \rangle \langle 12 \rangle\}$, through looking at the four-particle amplitude $A_4(1^{-1},2^{-1},3^{+1},4^{+1})$. The numerator must be $\langle 12 \rangle^2 [34]^2$. So, recognizing that $u = -s - t$ and $\tilde{u} = -\tilde{s} - \tilde{t}$, we see if we shift 
\begin{align}
& s = \langle 21 \rangle [12] \rightarrow \langle 21 \rangle ([12] + z \,  \tilde{s}/\langle 21 \rangle) = s + z \tilde{s} \\
& t =  \langle 41 \rangle [14] \rightarrow (\langle 41 \rangle + z \, \tilde{t}/[ 14 ])  [14]  = t +  z \tilde{t} \label{Validate1} \\
& u = -s -t \rightarrow u + z \, \tilde{u} = -(s+t) - z (\tilde{s} + \tilde{t}) \, . 
\end{align}
Deforming the anti-holomorphic part of $s$ and the holomorphic part of $t$ allows the $z$-shift to probe the $s$-, $t$-, and $u$-poles of the amplitude while leaving the numerator $\langle 12 \rangle^2 [34]^2$ unshifted.

This is the general case for amplitudes with higher-spin poles, i.e. for amplitudes with $3 \geq 2H + 1 - A \geq 2$ (the only cases amenable to this general analysis); we prove this by contradiction. By virtue of having two or three poles in each term, we are guaranteed that the numerator does not have any complete factors of $s$, $t$, and/or $u$: if it did, then this would knock out one of the poles in a term, in violation of the assumption that $N_p = 2$ or $3$.

\end{document}